\documentclass[twocolumn]{revtex4-1}
\usepackage[pdftex,letterpaper]{geometry}
\usepackage[pdftex]{color}
\usepackage{amsmath}
\usepackage{amssymb}
\usepackage[pdftex]{graphicx}

\makeatletter

\providecommand{\tabularnewline}{\\}

\usepackage{lineno}

\usepackage{xcolor}\usepackage{graphicx}\usepackage{xspace}\usepackage{colortbl}\usepackage{rotating}%
\usepackage[raggedrightboxes]{ragged2e}
\usepackage{textcomp}
\graphicspath{{ProblemSet_7-draft_graphics/}{ProblemSet_7-draft_tcache/}{ProblemSet_7-draft_gcache/}}
\DeclareGraphicsExtensions{.pdf,.eps,.ps,.png,.jpg,.jpeg}

\definecolor{tcGreen}{rgb}{0.1,0.5,0.1}

\makeatother

\begin{document}

\title{First Measurement of the 
Charged Current $\overline{\nu}_{\mu}$ 
Double Differential 
Cross Section on a Water Target without Pions in the final state}


\newcommand{\INSTHD}{\affiliation{University Autonoma Madrid, Department of Theoretical Physics, 28049 Madrid, Spain}}
\newcommand{\INSTEE}{\affiliation{University of Bern, Albert Einstein Center for Fundamental Physics, Laboratory for High Energy Physics (LHEP), Bern, Switzerland}}
\newcommand{\INSTFE}{\affiliation{Boston University, Department of Physics, Boston, Massachusetts, U.S.A.}}
\newcommand{\INSTD}{\affiliation{University of British Columbia, Department of Physics and Astronomy, Vancouver, British Columbia, Canada}}
\newcommand{\INSTGA}{\affiliation{University of California, Irvine, Department of Physics and Astronomy, Irvine, California, U.S.A.}}
\newcommand{\INSTI}{\affiliation{IRFU, CEA Saclay, Gif-sur-Yvette, France}}
\newcommand{\INSTGB}{\affiliation{University of Colorado at Boulder, Department of Physics, Boulder, Colorado, U.S.A.}}
\newcommand{\INSTFG}{\affiliation{Colorado State University, Department of Physics, Fort Collins, Colorado, U.S.A.}}
\newcommand{\INSTFH}{\affiliation{Duke University, Department of Physics, Durham, North Carolina, U.S.A.}}
\newcommand{\INSTBA}{\affiliation{Ecole Polytechnique, IN2P3-CNRS, Laboratoire Leprince-Ringuet, Palaiseau, France }}
\newcommand{\INSTEF}{\affiliation{ETH Zurich, Institute for Particle Physics, Zurich, Switzerland}}
\newcommand{\INSTIE}{\affiliation{CERN European Organization for Nuclear Research, CH-1211 Genève 23, Switzerland}}
\newcommand{\INSTEG}{\affiliation{University of Geneva, Section de Physique, DPNC, Geneva, Switzerland}}
\newcommand{\INSTHJ}{\affiliation{University of Glasgow, School of Physics and Astronomy, Glasgow, United Kingdom}}
\newcommand{\INSTDG}{\affiliation{H. Niewodniczanski Institute of Nuclear Physics PAN, Cracow, Poland}}
\newcommand{\INSTCB}{\affiliation{High Energy Accelerator Research Organization (KEK), Tsukuba, Ibaraki, Japan}}
\newcommand{\INSTIB}{\affiliation{University of Houston, Department of Physics, Houston, Texas, U.S.A.}}
\newcommand{\INSTED}{\affiliation{Institut de Fisica d'Altes Energies (IFAE), The Barcelona Institute of Science and Technology, Campus UAB, Bellaterra (Barcelona) Spain}}
\newcommand{\INSTEC}{\affiliation{IFIC (CSIC \& University of Valencia), Valencia, Spain}}
\newcommand{\INSTHH}{\affiliation{Institute For Interdisciplinary Research in Science and Education (IFIRSE), ICISE, Quy Nhon, Vietnam}}
\newcommand{\INSTEI}{\affiliation{Imperial College London, Department of Physics, London, United Kingdom}}
\newcommand{\INSTGF}{\affiliation{INFN Sezione di Bari and Universit\`a e Politecnico di Bari, Dipartimento Interuniversitario di Fisica, Bari, Italy}}
\newcommand{\INSTBE}{\affiliation{INFN Sezione di Napoli and Universit\`a di Napoli, Dipartimento di Fisica, Napoli, Italy}}
\newcommand{\INSTBF}{\affiliation{INFN Sezione di Padova and Universit\`a di Padova, Dipartimento di Fisica, Padova, Italy}}
\newcommand{\INSTBD}{\affiliation{INFN Sezione di Roma and Universit\`a di Roma ``La Sapienza'', Roma, Italy}}
\newcommand{\INSTEB}{\affiliation{Institute for Nuclear Research of the Russian Academy of Sciences, Moscow, Russia}}
\newcommand{\INSTHI}{\affiliation{International Centre of Physics, Institute of Physics (IOP), Vietnam Academy of Science and Technology (VAST), 10 Dao Tan, Ba Dinh, Hanoi, Vietnam}}
\newcommand{\INSTHA}{\affiliation{Kavli Institute for the Physics and Mathematics of the Universe (WPI), The University of Tokyo Institutes for Advanced Study, University of Tokyo, Kashiwa, Chiba, Japan}}
\newcommand{\INSTID}{\affiliation{Keio University, Department of Physics, Kanagawa, Japan}}
\newcommand{\INSTIF}{\affiliation{King's College London, Department of Physics, Strand, London WC2R 2LS, United Kingdom}}
\newcommand{\INSTCC}{\affiliation{Kobe University, Kobe, Japan}}
\newcommand{\INSTCD}{\affiliation{Kyoto University, Department of Physics, Kyoto, Japan}}
\newcommand{\INSTEJ}{\affiliation{Lancaster University, Physics Department, Lancaster, United Kingdom}}
\newcommand{\INSTFC}{\affiliation{University of Liverpool, Department of Physics, Liverpool, United Kingdom}}
\newcommand{\INSTFI}{\affiliation{Louisiana State University, Department of Physics and Astronomy, Baton Rouge, Louisiana, U.S.A.}}
\newcommand{\INSTHB}{\affiliation{Michigan State University, Department of Physics and Astronomy,  East Lansing, Michigan, U.S.A.}}
\newcommand{\INSTCE}{\affiliation{Miyagi University of Education, Department of Physics, Sendai, Japan}}
\newcommand{\INSTDF}{\affiliation{National Centre for Nuclear Research, Warsaw, Poland}}
\newcommand{\INSTFJ}{\affiliation{State University of New York at Stony Brook, Department of Physics and Astronomy, Stony Brook, New York, U.S.A.}}
\newcommand{\INSTGJ}{\affiliation{Okayama University, Department of Physics, Okayama, Japan}}
\newcommand{\INSTCF}{\affiliation{Osaka City University, Department of Physics, Osaka, Japan}}
\newcommand{\INSTGG}{\affiliation{Oxford University, Department of Physics, Oxford, United Kingdom}}
\newcommand{\INSTGC}{\affiliation{University of Pittsburgh, Department of Physics and Astronomy, Pittsburgh, Pennsylvania, U.S.A.}}
\newcommand{\INSTFA}{\affiliation{Queen Mary University of London, School of Physics and Astronomy, London, United Kingdom}}
\newcommand{\INSTE}{\affiliation{University of Regina, Department of Physics, Regina, Saskatchewan, Canada}}
\newcommand{\INSTGD}{\affiliation{University of Rochester, Department of Physics and Astronomy, Rochester, New York, U.S.A.}}
\newcommand{\INSTHC}{\affiliation{Royal Holloway University of London, Department of Physics, Egham, Surrey, United Kingdom}}
\newcommand{\INSTBC}{\affiliation{RWTH Aachen University, III. Physikalisches Institut, Aachen, Germany}}
\newcommand{\INSTFB}{\affiliation{University of Sheffield, Department of Physics and Astronomy, Sheffield, United Kingdom}}
\newcommand{\INSTDI}{\affiliation{University of Silesia, Institute of Physics, Katowice, Poland}}
\newcommand{\INSTIA}{\affiliation{SLAC National Accelerator Laboratory, Stanford University, Menlo Park, California, USA}}
\newcommand{\INSTBB}{\affiliation{Sorbonne Universit\'e, Universit\'e Paris Diderot, CNRS/IN2P3, Laboratoire de Physique Nucl\'eaire et de Hautes Energies (LPNHE), Paris, France}}
\newcommand{\INSTEH}{\affiliation{STFC, Rutherford Appleton Laboratory, Harwell Oxford,  and  Daresbury Laboratory, Warrington, United Kingdom}}
\newcommand{\INSTCH}{\affiliation{University of Tokyo, Department of Physics, Tokyo, Japan}}
\newcommand{\INSTBJ}{\affiliation{University of Tokyo, Institute for Cosmic Ray Research, Kamioka Observatory, Kamioka, Japan}}
\newcommand{\INSTCG}{\affiliation{University of Tokyo, Institute for Cosmic Ray Research, Research Center for Cosmic Neutrinos, Kashiwa, Japan}}
\newcommand{\INSTHF}{\affiliation{Tokyo Institute of Technology, Department of Physics, Tokyo, Japan}}
\newcommand{\INSTGI}{\affiliation{Tokyo Metropolitan University, Department of Physics, Tokyo, Japan}}
\newcommand{\INSTHG}{\affiliation{Tokyo University of Science, Faculty of Science and Technology, Department of Physics, Noda, Chiba, Japan}}
\newcommand{\INSTF}{\affiliation{University of Toronto, Department of Physics, Toronto, Ontario, Canada}}
\newcommand{\INSTB}{\affiliation{TRIUMF, Vancouver, British Columbia, Canada}}
\newcommand{\INSTG}{\affiliation{University of Victoria, Department of Physics and Astronomy, Victoria, British Columbia, Canada}}
\newcommand{\INSTDJ}{\affiliation{University of Warsaw, Faculty of Physics, Warsaw, Poland}}
\newcommand{\INSTDH}{\affiliation{Warsaw University of Technology, Institute of Radioelectronics and Multimedia Technology, Warsaw, Poland}}
\newcommand{\INSTFD}{\affiliation{University of Warwick, Department of Physics, Coventry, United Kingdom}}
\newcommand{\INSTGH}{\affiliation{University of Winnipeg, Department of Physics, Winnipeg, Manitoba, Canada}}
\newcommand{\INSTEA}{\affiliation{Wroclaw University, Faculty of Physics and Astronomy, Wroclaw, Poland}}
\newcommand{\INSTHE}{\affiliation{Yokohama National University, Faculty of Engineering, Yokohama, Japan}}
\newcommand{\INSTH}{\affiliation{York University, Department of Physics and Astronomy, Toronto, Ontario, Canada}}

\INSTHD
\INSTEE
\INSTFE
\INSTD
\INSTGA
\INSTI
\INSTGB
\INSTFG
\INSTFH
\INSTBA
\INSTEF
\INSTIE
\INSTEG
\INSTHJ
\INSTDG
\INSTCB
\INSTIB
\INSTED
\INSTEC
\INSTHH
\INSTEI
\INSTGF
\INSTBE
\INSTBF
\INSTBD
\INSTEB
\INSTHI
\INSTHA
\INSTID
\INSTIF
\INSTCC
\INSTCD
\INSTEJ
\INSTFC
\INSTFI
\INSTHB
\INSTCE
\INSTDF
\INSTFJ
\INSTGJ
\INSTCF
\INSTGG
\INSTGC
\INSTFA
\INSTE
\INSTGD
\INSTHC
\INSTBC
\INSTFB
\INSTDI
\INSTIA
\INSTBB
\INSTEH
\INSTCH
\INSTBJ
\INSTCG
\INSTHF
\INSTGI
\INSTHG
\INSTF
\INSTB
\INSTG
\INSTDJ
\INSTDH
\INSTFD
\INSTGH
\INSTEA
\INSTHE
\INSTH

\author{K.\,Abe}\INSTBJ
\author{R.\,Akutsu}\INSTCG
\author{A.\,Ali}\INSTCD
\author{C.\,Alt}\INSTEF
\author{C.\,Andreopoulos}\INSTEH\INSTFC
\author{L.\,Anthony}\INSTFC
\author{M.\,Antonova}\INSTEC
\author{S.\,Aoki}\INSTCC
\author{A.\,Ariga}\INSTEE
\author{Y.\,Ashida}\INSTCD
% ...

\collaboration{The T2K Collaboration}\noaffiliation
\begin{abstract}
This paper reports the first differential measurement of the charged-current
 $\overline{\nu}_{\mu}$
interaction cross section on water with no pions in the final state. 
The unfolded flux-averaged measurement
using the T2K off-axis near detector is given in double differential
bins of $\mu^+$ momentum and angle.
The integrated cross section in a restricted phase space 
is $\sigma=\left(1.11\pm0.18\right)\times10^{-38}$ cm$^{2}$ per
water molecule. Comparisons with several nuclear models are also presented.

\end{abstract}
\maketitle

\section{Introduction}

Long baseline neutrino experiments \cite{Abe:2018_CPV-T2K_b,novaOA2017}
are now measuring both neutrino ($\nu_{\mu}\rightarrow\nu_{e}$) and
antineutrino ($\overline{\nu}_{\mu}\rightarrow\overline{\nu}_{e}$)
appearance oscillations to determine fundamental neutrino mixing parameters
and
to search for charge-parity (CP) violation
in the lepton sector. Testing this symmetry may answer one of the
most fundamental physics questions, the mystery of the matter-antimatter
imbalance in our Universe.

Neutrino oscillation measurements are performed by measuring neutrino interactions on nuclei.
The present uncertainties on models describing the (anti)neutrino-nucleus scattering are the main 
source of systematic error in currently operating
experiments, such as T2K \cite{Abe:2011ks} and NOvA \cite{novafirst}, and will affect future 
projects, DUNE \cite{Acciarri:2015uup} and HyperKamiokande \cite{Abe:2014oxa}. 
The main difficulty in the description of (anti)neutrino-nucleus interactions derives from the intrinsic nature of
the nucleus, where nucleons are bound together and nuclear effects must be taken into account. Many models are currently available,
describing different pieces of this complex scenario such as relativistic Fermi gas \cite{smith-moniz}, Spectral Function
\cite{Benhar:1995, Ankowski:2010yh},
the random phase approximation \cite{Singh:1992dc, Gil:1997bm, Nieves:2004wx, Nieves:2005rq}, and the multinucleon
\cite{2P2H-Nieves, Martini:2009uj, Delorme:1985ps, Marteau:1999jp, Martini:2010ex, Nieves:2011yp, Martini:2011wp, Nieves:2012,
Buss:2011mx, Gallmeister:2016dnq, susav2}
models. 
Thus a key component required by present and
future \cite{Acciarri:2015uup,Abe:2015zbg} experiments are the precise measurements
and tests of theoretical models of both neutrino and antineutrino
cross sections on detector target materials such as scintillator,
water, and liquid Argon. In charged current interactions 
without pions in the final state, 
detailed measurements of the outgoing muon will help to test different theoretical models. In this paper, using
the off-axis near detector of the T2K experiment, we present the first
double differential antineutrino cross section measurement on water
and compare it to various model predictions.

Measurements by T2K probe the completeness of the interaction model 
by comparing neutrinos and antineutrinos \cite{Abe:2017csu}, by using
different target materials \cite{Abe:2016tmq}, \cite{Abe:2018tianlu}, 
and different energy spectra \cite{Abe-INGRID-2014a, Abe-INGRID-2016,Abe-INGRID-2019}, 
and through leptonic-hadronic state correlations\cite{Abe:2018dolan}.
The published T2K measurements used 
unfolding techniques such as the
 D'Agostini iterative unfolding \cite{Abe:2018tianlu} or the maximum binned
likelihood \cite{Abe:2016tmq,Abe:2018dolan}. 

The analysis in this paper
determines the kinematics of the
outgoing $\mu^+$ produced in  
$\bar\nu_\mu$ CC0$\pi$ interactions on water.
The differential cross sections are
extracted by following a similar analysis procedure performed
in a previous T2K publication \cite{Abe:2018dolan}. 

In the following sections, we describe the T2K anti-neutrino beam
and near detector (ND280), the Monte Carlo simulation and data samples,
the event selection, 
the cross section extraction method, 
the results
and model comparisons.

\section{T2K EXPERIMENT}

The Tokai to Kamioka (T2K) experiment \cite{Abe:2011ks} is a long baseline
neutrino experiment located in Japan. It is composed
of a neutrino beamline and a near detector at the Japan Proton Accelerator
Research Complex (J-PARC) laboratory in Tokai, and a far detector,
Super Kamiokande (SK), that is situated 295 km away in the 
Mozumi Mine in the Kamioka area of Hida City.
The J-PARC synchrotron produces a 30 GeV energy proton beam
that strikes a graphite target to produce pions and kaons that are
focused by three horn magnets into a 96 m long decay volume. The horn
magnet polarity can be set to select either positively or negatively charged pions
and kaons to produce a predominately neutrino or antineutrino beam.
The magnet setting for positively charged tracks is denoted as Forward Horn
Current (FHC) and for negatively charged tracks, Reverse Horn Current (RHC).
The near detector complex, 280 m downstream of the target, consists of an on-axis
detector (INGRID) and an off-axis detector (ND280). 
The ND280 and SK detectors are positioned 2.5$^{\circ}$ away from the neutrino beam axis. 
At this angle, neutrino and antineutrino beams energies peak near 0.6 GeV.
The following subsections describe the $\bar\nu_\mu$
beam, the ND280 detector, and the Monte Carlo simulation programs.

\subsection{T2K BEAM}

\begin{figure}
\includegraphics[scale=1.2]{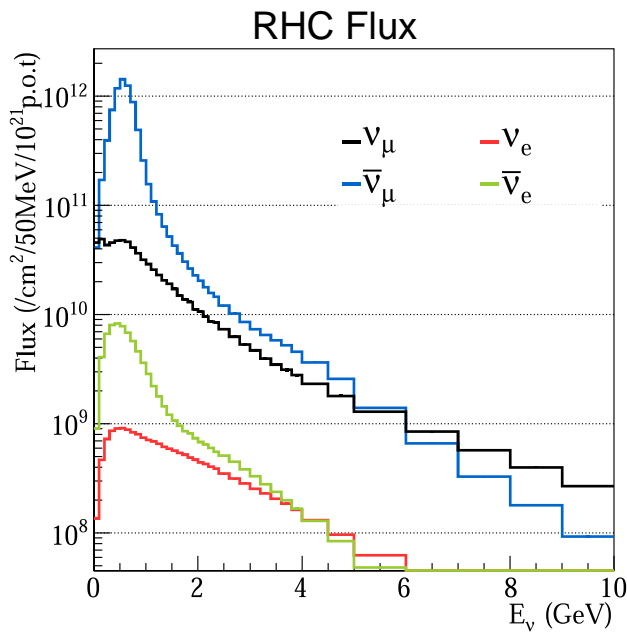}

\caption{\label{fig:RHC-flux}The RHC flux given per cm$^{2}$/50 MeV/$10^{21}$ Protons on Target (PoT) as a function of
energy at the ND280 detector for the different neutrino components ($\overline{\nu}_{\mu}$, $\nu_{\mu}$, $\overline{\nu}_{e}$, $\nu_e$). }
\end{figure}

The neutrino and antineutrino fluxes for the RHC configuration in the ND280 detector
were determined by simulating the T2K neutrino beamline \cite{t2kflux}
using FLUKA2011 \cite{Fluka:2014x,Ferrari:2005zk}, GEANT3 \cite{GEANT3},
and GCALOR \cite{GCALOR} software packages. The simulated hadronic
yields have been reweighted using the NA61/SHINE \cite{Abgrall:2011ae,Abgrall:2011ts,Abgrall:2016fs}
thin-target measurements and 
this reduced the 
flux uncertainties to be less than 10\% around the flux
peak.
The $\overline{\nu}_{\mu}$ fluxes are plotted in Fig. \ref{fig:RHC-flux}
along with the three background neutrino flavors, $\nu_{\mu}$, $\nu_{e}$,
and $\bar{\nu}_{e}$. In the peak region ($\sim0.6$
GeV ) the $\nu_{\mu}$ contamination in the antineutrino flux is $\sim2.5\%$.
Details on the
antineutrino beam and comparisons to the neutrino beam have been discussed
in a previous T2K publication \cite{Abe:2017vif}.

\subsection{ND280 DETECTOR.}

The ND280 detector consists of sub-detectors inside the refurbished
UA1/NOMAD magnet that operates at a 0.2 T magnetic field, that is normal
to the neutrino beam and the vertical direction. The ND280 sub-detectors
include the $\pi^{0}$ detector \cite{Assylbekov:2011sh} (P$\emptyset$D),
three tracking time projection chambers \cite{Abgrall:2010hi} (TPC1-3),
two fine-grained detectors (FGD1-2) interleaved with TPC1-3, and an
electromagnetic calorimeter (ECAL), that encloses the P$\emptyset$D,
TPC1-3, and FGD1-2 sub-detectors. For the analysis reported in this paper,
the P$\emptyset$D and the TPC tracking detector in the ND280 detector
complex are used. We define the +Z direction parallel to the neutrino
beam direction, and +Y direction 
pointing vertically upwards. 

\begin{figure}
\includegraphics[scale=0.38]{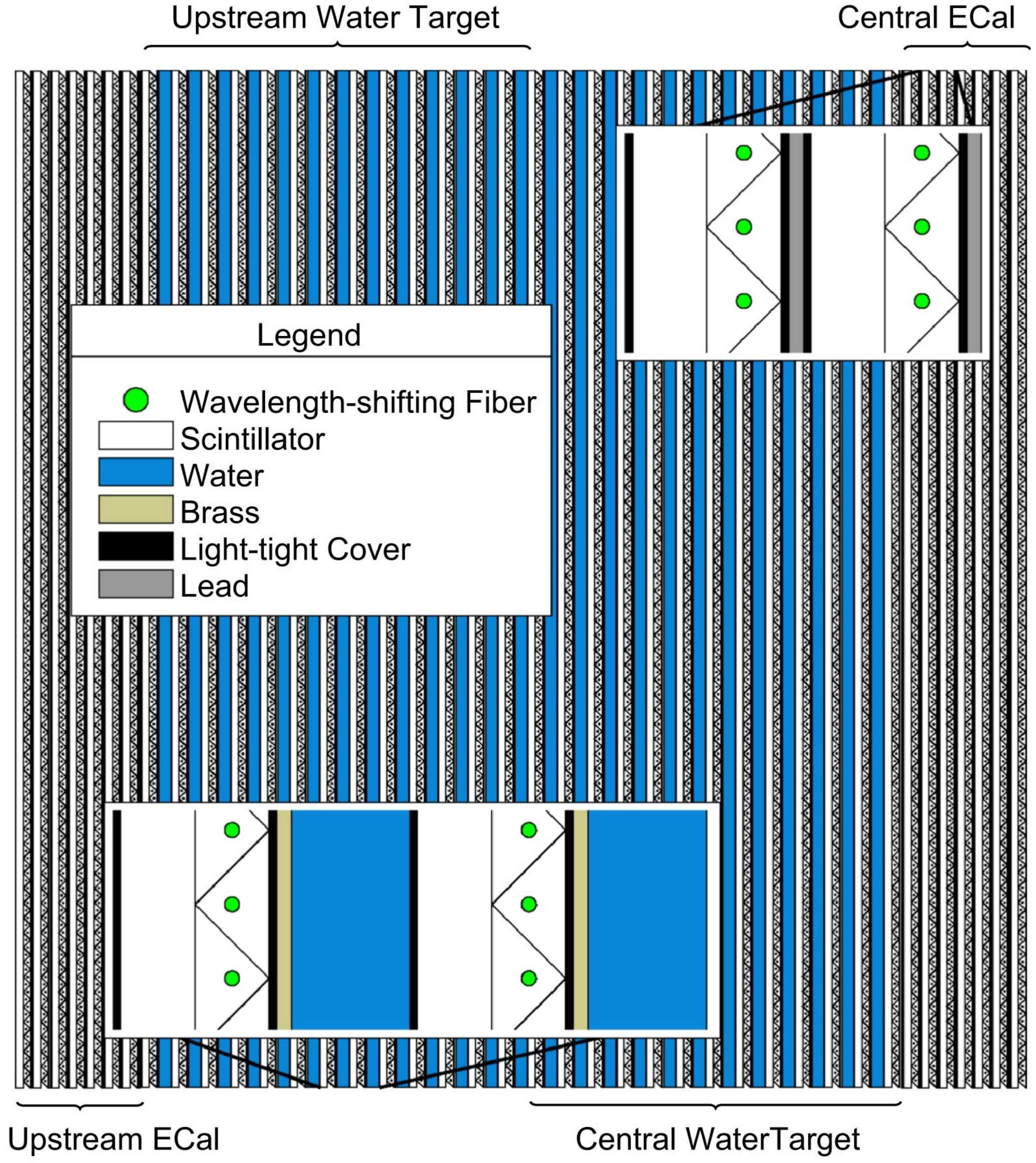}

\caption{\label{fig:Side-view-schematic}Side view schematic diagram of the
P$\emptyset$D detector. The white, zig-zag, and blue strip regions
represent the vertical scintillator bars, the horizontal scintillator
bars, and the water bag regions, respectively. The vertical and horizontal
bars represent an x-y module or P$\emptyset$Dule. The first and last
groups of seven P$\emptyset$Dules form the upstream and the central
ECAL ``super'' modules and the middle 26 P$\emptyset$Dules interleaved
with the water bags are the water target region. In this drawing, the beam direction (+Z) is to the
right, the +Y direction is up, and +X direction is into the drawing.}
\end{figure}

We describe detector details relevant for the analysis. 
The P$\emptyset$D detector that reconstructs the neutrino interaction
vertex is shown in Fig. \ref{fig:Side-view-schematic}. 
It contains 40 scintillator module planes (called P$\emptyset$Dules), each built of two 
perpendicular arrays of triangular scintillator bars, 134 horizontal (X) and 126 
vertical (Y) bars. Each bar has a wavelength shifting fiber centered in the bar that is read out by a Hamamatsu 
Multi-pixel photon counter. P0Dules are formed into 3 major groups.
The center group, called the water target, is the primary target for this analysis.
It has 26 P$\emptyset$Dules interleaved with 2.8 cm thick water bags and 1.3 mm thick brass sheets.
The water target region is drainable and data can be taken with or without water.
The fiducial volume mass is 1900 kg of water and 3570 kg of other materials.
The two other regions (called upstream and central ECALs) are the upstream and downstream groups that each 
contain 7 P$\emptyset$Dules sandwiched with lead sheets clad with steel.
These two groups form a veto region to isolate neutrino interactions that occur in the water target.
The size of the entire active P0D volume is $2103\times2239\times2400$ mm$^{3}$ 
(XYZ) and its mass with 
and without water is 15,800 kg and 12,900 kg respectively.
The two other regions (called
upstream and central ECALs) are the upstream and downstream groups
that each contain 7 P$\emptyset$Dules and steel sheets clad with
lead. 
These two groups form a veto region to isolate neutrino
interactions that occur in the water target. 

The charged current neutrino interaction in the P$\emptyset$D, creates
a muon that exits the P$\emptyset$D and enters the TPC1-3 detectors.
The TPC1-3 detectors measure the $\mu^+$ momentum and its dE/dx energy
loss which is used for muon particle identification.

\section{Data and Monte Carlo Samples}

The studies reported here used the
RHC  $\overline{\nu}_{\mu}$ beam running mode. 
The runs utilized
detector configurations where the P$\emptyset$D water bags were
filled (water-in) or empty (water-out).  Roughly equal amounts
of exposure in each configuration was allowed in each running period so that the detector operations, efficiencies,
and beam conditions were similar for both the water-in
and water-out data samples.

\begin{table}
\begin{tabular}{ccc}
\hline 
P$\emptyset$D Target  & Data & MC\tabularnewline
Mode & sample & sample\tabularnewline
\hline 
\hline 
water-in & $2.87\times10^{20}$  &  $20.8\times10^{20}$ \tabularnewline
water-out & $3.43\times10^{20}$  &  $20.9\times10^{20}$\tabularnewline
\hline 
\end{tabular}

\caption{\label{tab:data-mc-samples}Protons on Target (PoT) for data and equivalent
MC samples for RHC antineutrino beam running split for P$\emptyset$D
water-in/water-out modes.}
\end{table}

\subsection{Data Samples}

The total 
Proton on Target (PoT) exposure
for RHC antineutrino beam data running is shown in Table I.
This sample required all data quality cuts to be satisfied and 
corresponded to $2.87\times10^{20}$
PoT for the water-in and $3.43\times10^{20}$ PoT for the water-out
modes. 

\subsection{Monte Carlo Simulation}

The analysis utilized simulated Monte Carlo (MC) samples with different
beam and detector configurations for each data run. The total MC combined
water-in and out samples were equivalent to $20.8\times10^{20}$ and
$20.9\times10^{20}$ PoT, respectively. The simulation includes: 
\begin{enumerate}
\item Primary $\bar\nu_\mu$ and background $\nu_\mu$, $\nu_e$, and $\bar\nu_e$  beam production in the
graphite target and propagation through the following horns and decay
volume. 
The hadronic rates from the beam target were generated by FLUKA2011 
which was tuned to the NA61/SHINE measurements and the GEANT3
simulation software predicted the flux and energy spectrum
for the different neutrino flavors.  
\item The antineutrino and neutrino interactions in the ND280 detector,
where the NEUT \cite{Hayato:2009} MC generator (version 5.3.3) is
used to calculate the interaction cross sections and the final state
particle kinematics. 
\item The detector response, which used the GEANT4 \cite{Agostinelli:2002hh}
simulation package to
transport the final state particles through the ND280 detector complex. 
\end{enumerate}

\section{Event and Kinematic Selection}

The event selection for antineutrino interactions is optimized to
identify the observable charged current events with no charged or neutral pions in the
final state. This is nominally denoted as the CC-$0\pi$ final state. This mainly
includes charged current quasi-elastic (CCQE) events and the case
where pions are created in the primary resonant 
antineutrino
interaction, but reabsorbed before exiting the nucleus. The 
$\bar\nu_\mu$ interactions with a multi-nucleon state such as 2 particle-2 hole (2p2h)
can produce a final state without mesons. Non-CCQE neutrino
interactions that produce a CC-0$\pi$ final state will have
antineutrino kinematics that are different from those created in CCQE interactions.
This will be important to understand and to carefully model since this can change the
antineutrino energy reconstruction which can affect current and future
neutrino oscillation analyses.


We first consider three antineutrino mode selections (CC-inc, CC-0$\pi$, and CC-1$\pi$).
The event selection is similar to a previous T2K analysis \cite{Abe:2018tianlu}
of a neutrino differential cross section measurement on water in the P$\emptyset$D
detector. The selection requires:
\begin{enumerate}
\item Overall ND280 data quality flags are good such that the detector was
operational and stable during taking data.
\item There is a reconstructed track in the P$\emptyset$D matching a track
in the TPC with the start of the track reconstructed in the fiducial
volume of the P$\emptyset$D water target.
\item There is at least one track reconstructed in TPC1
\item There is a muon track candidate that is the highest momentum positively charged track,
the highest momentum track in the event, 
and has a TPC dE/dx track measurement consistent
		with a muon energy loss.  These first four requirements
define the CC-Inc event selection.
\item There are no reconstructed P$\emptyset$D showers in the event.
This cut removes charged current events with a $\pi^{0}$.
\item Remaining events are then separated into 3 categories based on the number of $\mu$-like
	  P$\emptyset$D tracks in the event.
\begin{enumerate}
	\item Events with only a muon track candidate define the CC-0$\pi$ selection.
	\item Events with a muon track candidate and one $\mu$-like track define the CC-1$\pi$ selection.
	\item All other remaining events are not selected.
\end{enumerate}
\end{enumerate}
If there are other tracks, besides the muon track candidate, they are defined as $\mu$-like 
if the average energy loss per P$\emptyset$D layer
near the middle of the track is less than 1.5
times that of the muon track candidate in the same event.
The $\mu^+$ track candidate is
a minimum ionizing particle track which should have nearly the same 
measured energy loss
per unit length of the pion track
as measured in between the interaction vertex and before it
decays in the detector.
Comparing the average energy losses between
the muon track candidate and
different P$\emptyset$D tracks in the same event, ensures that the tracks use
the same detector gain calibrations. 
Using this cut, proton and pion tracks can be differentiated, allowing for any number of protons to be 
present in CC-0$\pi$ events.

In Table \ref{tab:Purity-and-efficiency}, 
the purity and the efficiency of
the three selections (columns 2-4) are given in terms of five true
MC final states (column 1). The true final states are
CC-0$\pi$, CC-1$\pi$, CC-other (all other CC states
excluding  CC-0$\pi$ and CC-1$\pi$), 
BKGD (neutral current and non-$\bar\nu_\mu$ interactions) 
and OOFV (out of fiducial volume events). 
The OOFV events have interactions that occur outside the
selected P$\emptyset$D target region.
This table shows that the CC-$0\pi$ selection
has very good purity $(\sim80\%)$ and very high efficiency $(\sim95\%)$ relative to the CC-Inc sample. 

\begin{figure*}
\
\includegraphics[scale=0.42]{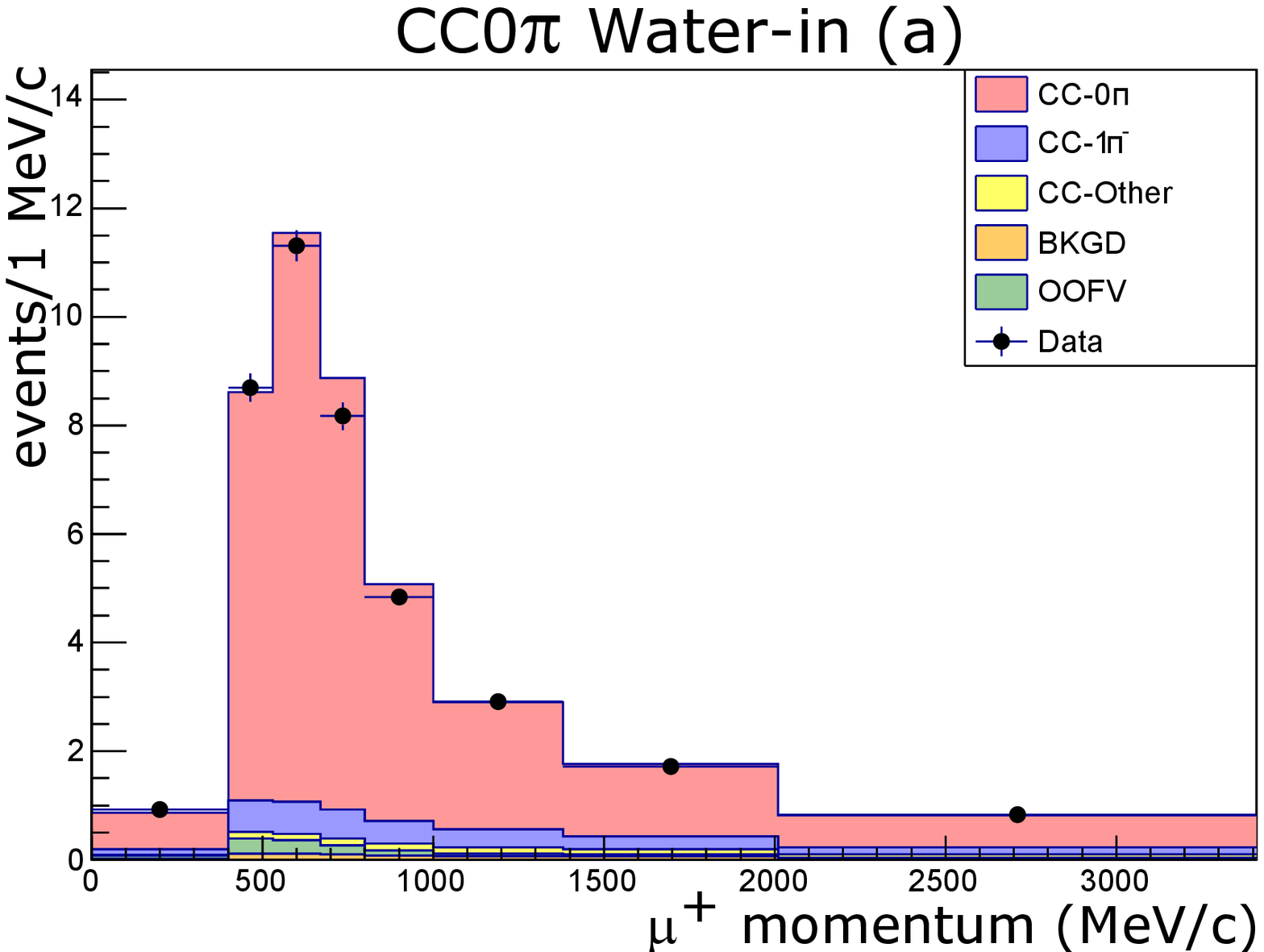}
\includegraphics[scale=0.42]{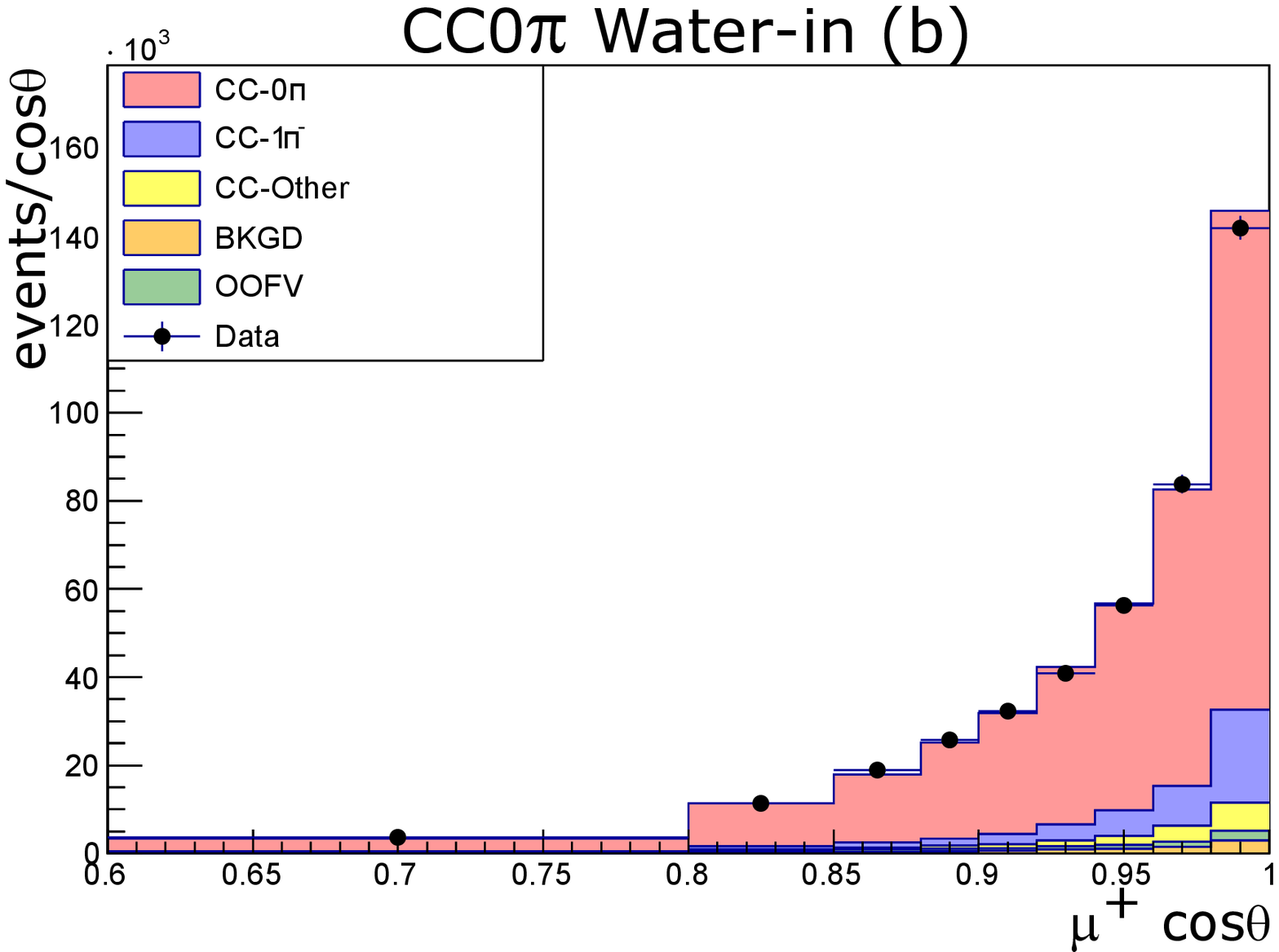}
\includegraphics[scale=0.42]{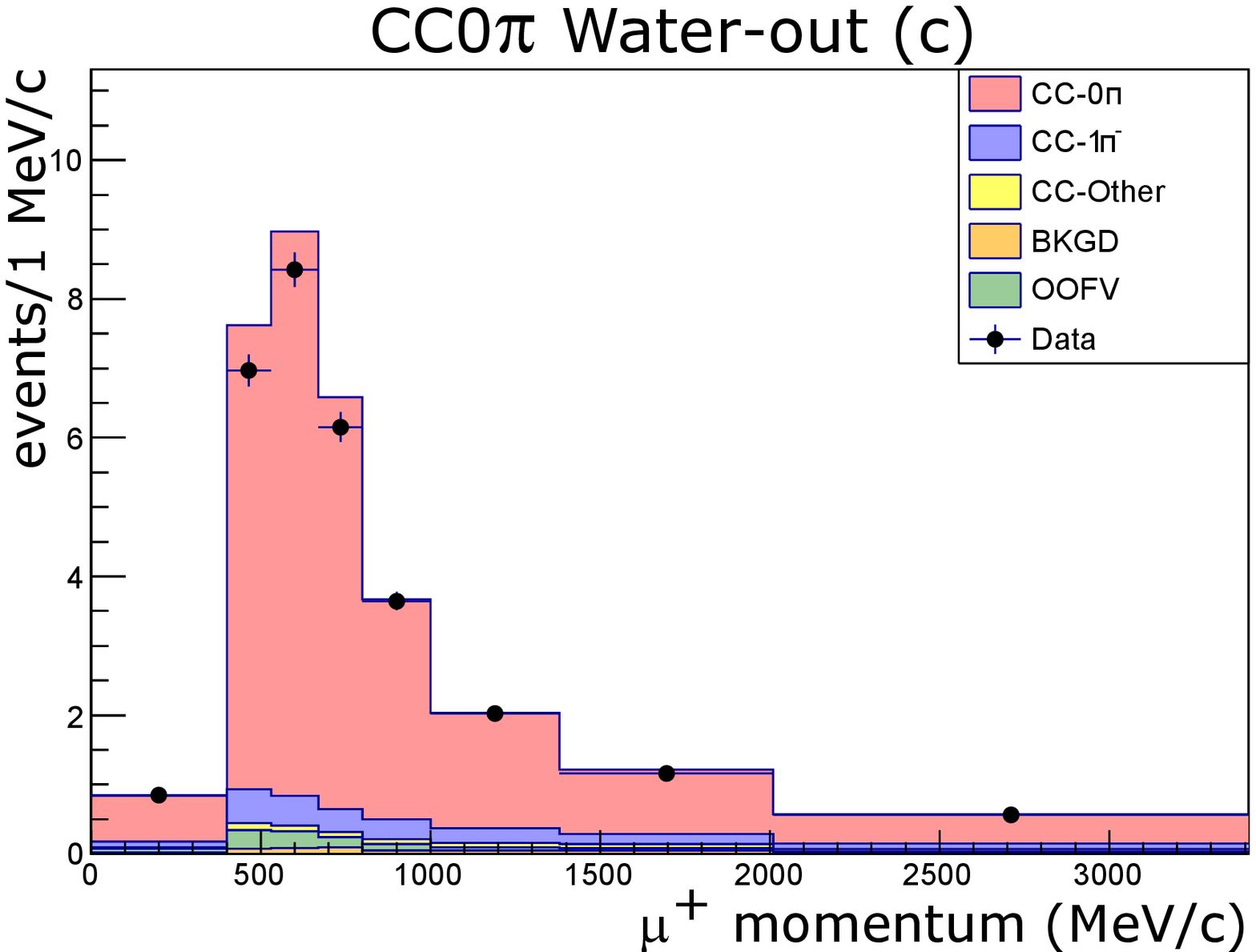}
\includegraphics[scale=0.42]{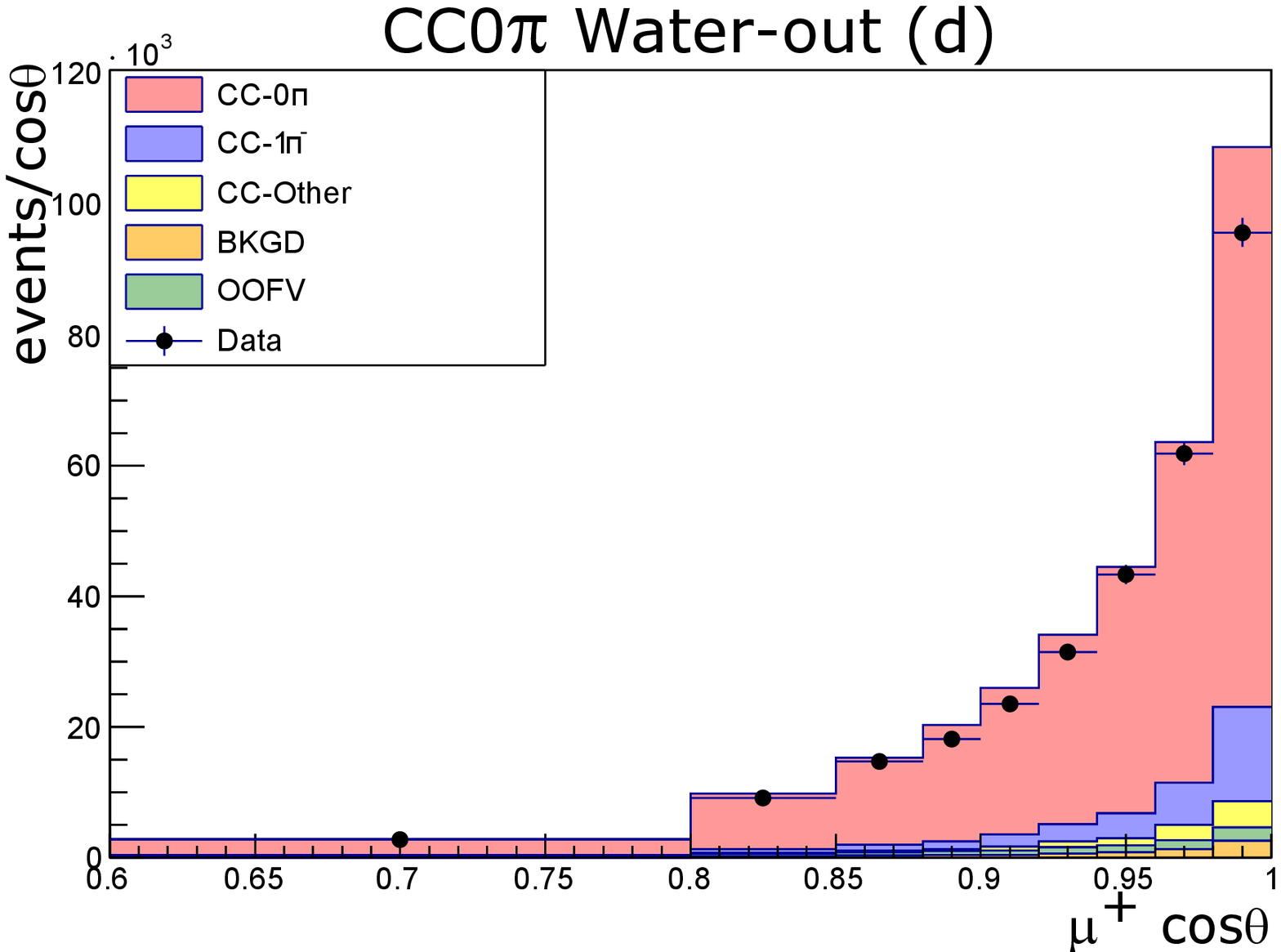}

\includegraphics[scale=0.42]{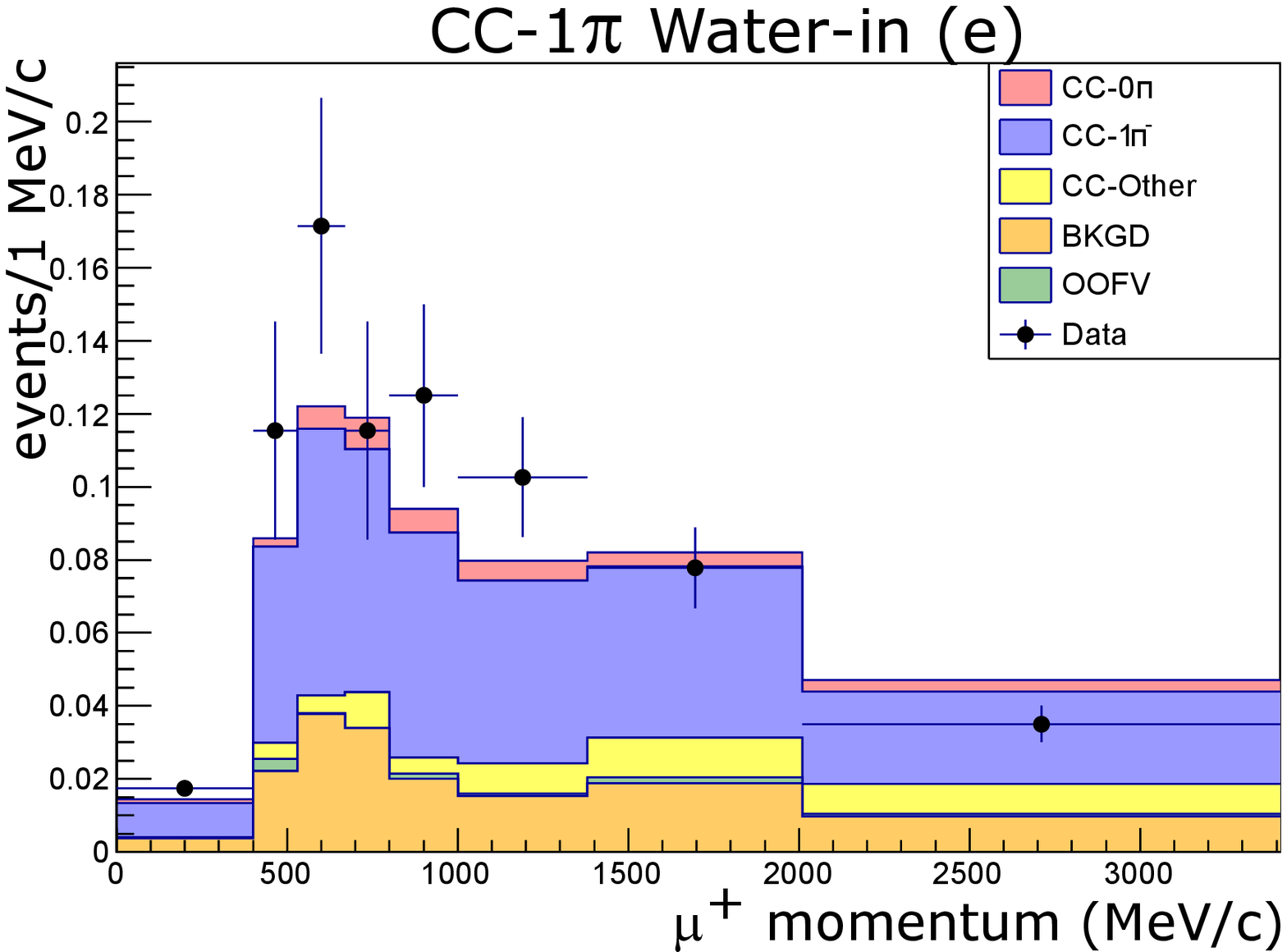}
\includegraphics[scale=0.42]{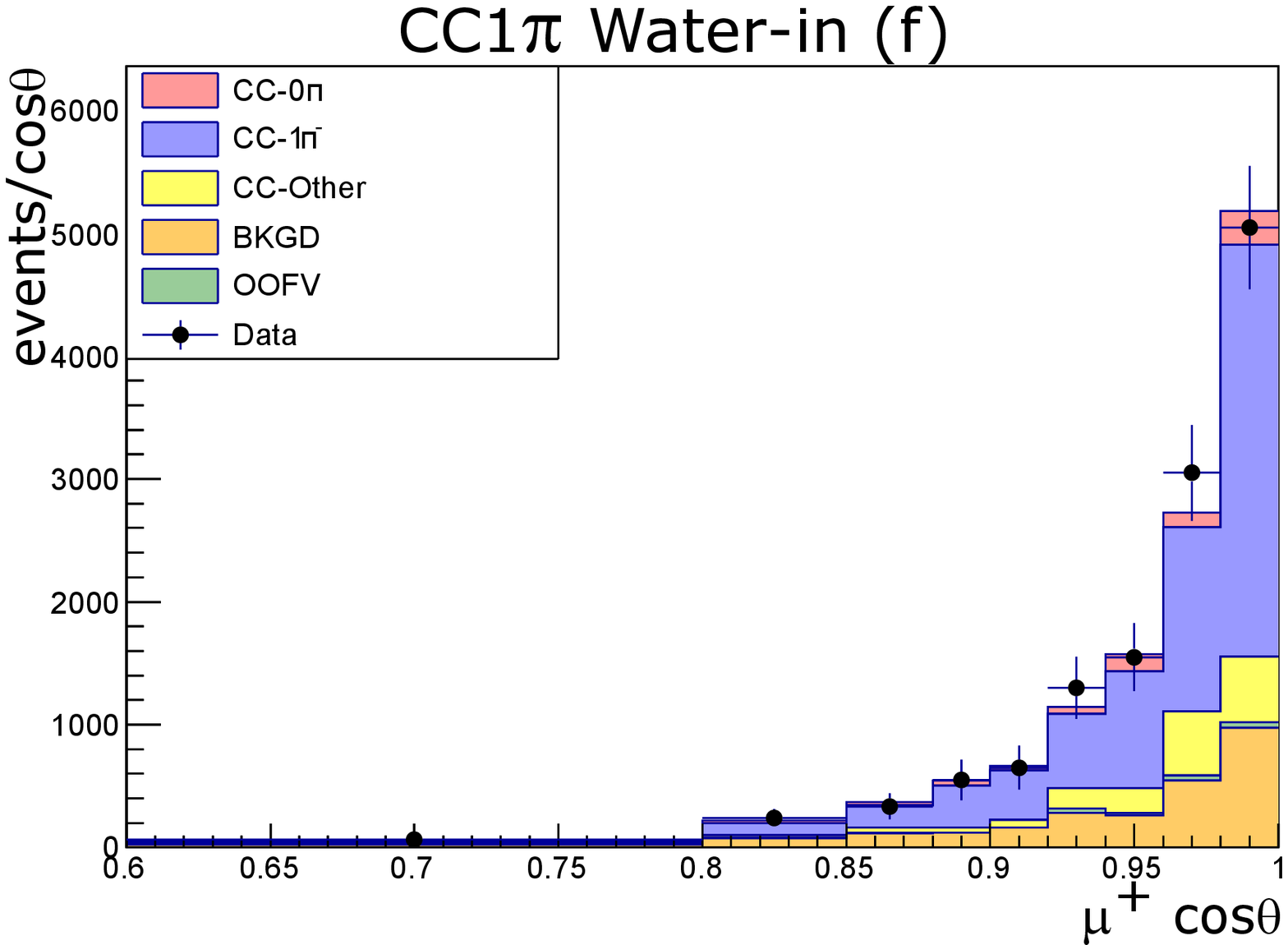}
\includegraphics[scale=0.42]{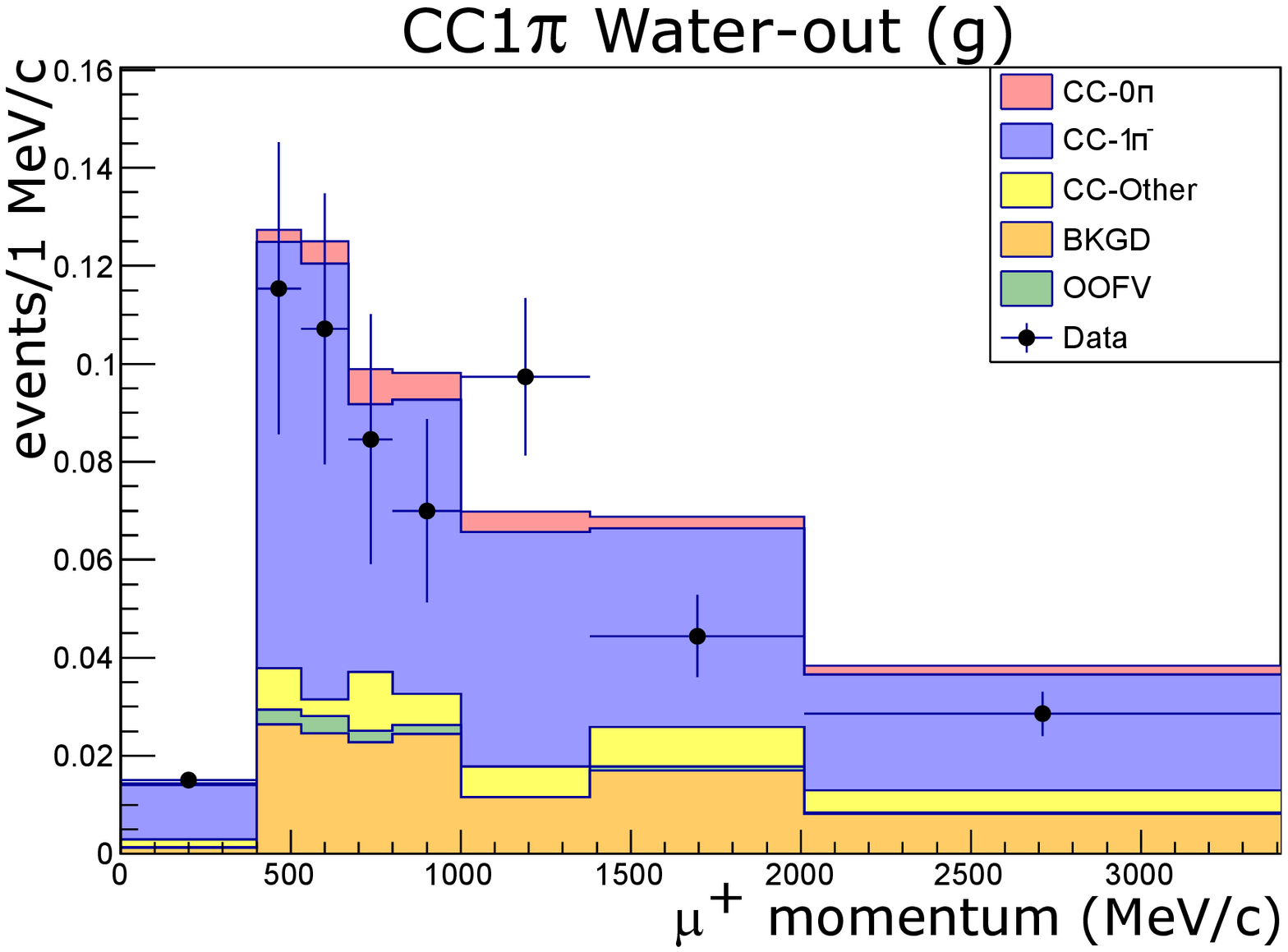}
\includegraphics[scale=0.42]{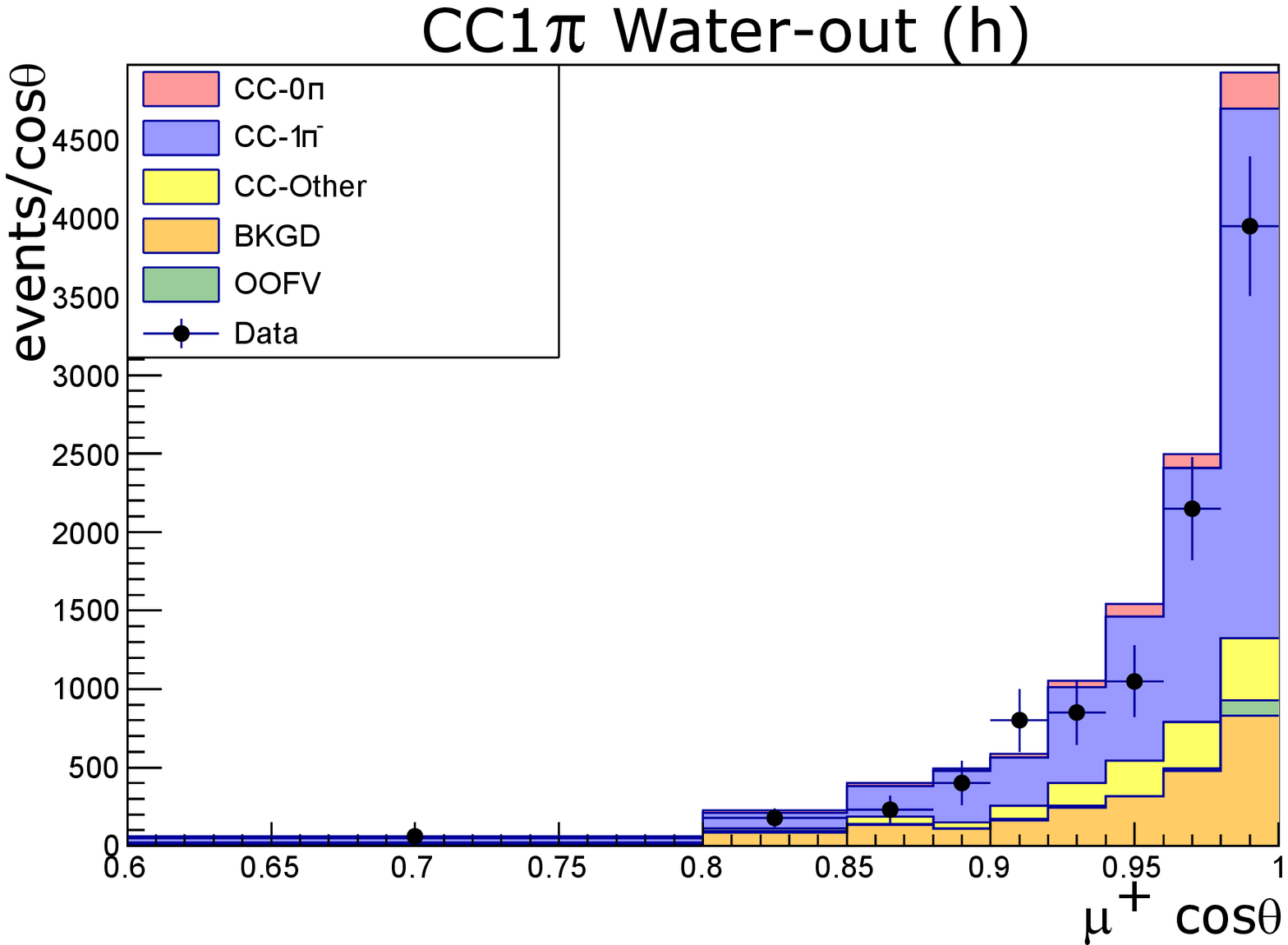}
\caption{\label{fig:Data+MC_mom+costht}The comparisons of 
lab frame momentum
(left column) and $\cos\theta$ (right column) distributions between
data (black dots with error bars)
and NEUT simulation predictions before fitting (stacked color bands).
The CC-$0\pi$ selections have been applied on the
water-in samples (top row, (a) and (b)) and water-out samples (second row, (c) and (.d))
The CC-$1\pi$ selections have been applied on the 
water-in samples (third row, (e) and (f)) and water-out samples (fourth row, (g) and (h)).}
\end{figure*}

In Fig. \ref{fig:Data+MC_mom+costht} are shown
the plots of the CC-0$\pi$ and CC-1$\pi$ selections of data superimposed
over the NEUT simulations. This is presented 
in pairs of water-in/out samples for the CC-$0\pi$ momentum,
the CC-$0\pi$ $\cos\theta$, the CC-$1\pi$ momentum,
and the CC-$1\pi$ $\cos\theta$. The Monte Carlo color
bands correspond to the true CC-$0\pi$, CC-$1\pi$,
CC-Other, BKGD, and OOFV events. 
Overall there is reasonable agreement between data and Monte Carlo. 

In Table II and Fig. 3 (a-d), the dominant backgrounds
for the CC-0$\pi$ selection are caused by misidentified CC events
with one emitted pion (CC-1$\pi$) or CC-other events, 
with CC-1$\pi$ being the largest of the two.
In order to constrain the CC-1$\pi$ background, a control sample of CC-1$\pi$ selected events
will be included in the analysis fitting described in the next section. This allows a data constraint on the background
estimation which leads to smaller background modeling uncertainties.

\begin{table}
\begin{tabular}{lccc}
\hline 
Water-in mode: &  & \% in Selected Sample & \tabularnewline
\hline 
\hline 
  & CC-Inc & CC-0$\pi$ & CC-1$\pi$\tabularnewline
\hline 
CC-0$\pi$ & 60 & 80 & 10\tabularnewline
\hline 
CC-1$\pi$ & 17 & 13 & 57\tabularnewline
\hline 
CC-Other & 13 & 3 & 15\tabularnewline
\hline 
BKGD & 7 & 1 & 15\tabularnewline
\hline 
OOFV & 4 & 2 & 3\tabularnewline
\hline 
$\epsilon_{relative}$ &  & 96 & 14\tabularnewline
\hline 
Water-out mode: &  & \% in Selected Sample & \tabularnewline
\hline 
\hline 
 & CC-Inc & CC-0$\pi$ & CC-1$\pi$\tabularnewline
\hline 
CC-0$\pi$ & 58 & 82 & 11\tabularnewline
\hline 
CC-1$\pi$ & 16 & 12 & 57\tabularnewline
\hline 
CC-other & 12 & 2 & 14\tabularnewline
\hline 
BKGD & 8 & 1 & 14\tabularnewline
\hline 
OOFV & 5 & 2 & 4\tabularnewline
\hline 
$\epsilon_{relative}$ &  & 95 & 15\tabularnewline
\hline 
\end{tabular}

\caption{\label{tab:Purity-and-efficiency}Purity and efficiency tables for
	the different selections for water-in and water-out samples. The true
	final states are given in the first column and the three selections (CC-Inc, CC-0$\pi$,
	and CC-1$\pi$) are given in the rows below the double lines. An example in this table
	is that the water-out mode CC-0$\pi$ selected sample will have 82\% of its event originate from
	the true CC-0$\pi$ final state.  The $\epsilon_{relative}$ is the fraction of relevant 
	events (CC-0$\pi$ or CC-1$\pi$) present in the 
	CC-Inc sample retained by the number of $\mu$-like tracks requirement.  For example, 96\% of the CC-0$\pi$ events 
	present in the water-in CC-Inc sample are retained in the water-in CC-0$\pi$ sample.
	See text for final state descriptions.
	}
\end{table}

\section{Double Differential Cross Section Fitting Method}

In this section we first describe the fitting and unfolding technique
to extract the differential cross section in true $p-\cos\theta$
bins 
of the $\mu^+$ track. 
Then the binning choice is explained followed by descriptions
of the fit parameters and checks and validation on the fitting method.
Finally, the regularization choice and overall checks are discussed.

\subsection{Fitting}

In an idealized experiment with no backgrounds and perfect detector resolutions, the differential
cross section as a function of kinematic variable $x$ in a particular
bin $\Delta x_{j}$ is given as: 
\begin{equation}
\frac{d\sigma}{dx_{j}}=\frac{N_{j}}{\epsilon_{j}\Phi T\Delta x_{j}}
\end{equation}
where $N_{j}$ is the number of measured events in bin $j$, $T$
is the number of target nuclei, $\Phi$ is the neutrino flux per unit
area and $\epsilon_{j}$ is the efficiency to reconstruct a signal
event in bin $j$. 
In this analysis, the $\Delta x_{j}$ is the
$p-\cos\theta$ bin of the $\mu^+$ track in the lab frame. 
We define $N_{j}^{sig}$ as the number signal events 
and $N_{j}^{sig,MC}$ as the number of predicted MC events in $p-\cos\theta$ bin $j$. 
We introduce a scale parameter, $c_{j}$, to be fitted, where:
\begin{equation}
N_{j}^{sig}=c_{j}N_{j}^{sig,MC}
\end{equation}

If we include different background types $k$ in the reconstructed
data, $\sum_{k}^{bkgd\ types}N_{j}^{bkgd\ k,MC}$ should be
added to the above equation. In addition, if the background event rates depend on
different model parameters, the backgrounds can be reweighted by
a product term, $\prod_{a}^{model}\omega\left(\vec{a}\right)_{j}^{k}$
which depends on a vector $\vec{a}$ of background model parameters. Then
the expression becomes:
\begin{equation}
N_{j}=c_{j}N_{j}^{sig,MC}+\sum_{k}^{bkgd\ types}\left(\prod_{a}^{model}\omega\left(\vec{a}\right)_{j}^{k}\right)N_{j}^{bkgd\ k,MC}
\end{equation}
where $N_{j}$ is the predicted number of measured events (signal+background)
in bin $j$, fitted parameters are $c_{j}$ and vector parameter
$\vec{a}$.

In real experiments the reconstruction is not perfect and we need
to allow for smearing where events from a particular true $p-\cos\theta$
bin $j$ were smeared over several different reconstructed $p-\cos\theta$
 bins.  If we consider events in some true kinematic bin $j$ that are reconstructed with kinematics across 
 bins indexed by $i$, a ``smearing matrix''
$S_{ij}$ can be constructed:
\begin{equation}
S_{ij}=\frac{N_{reco\ in\ i}^{true\ in\ j}}{N^{true\ in\ j}}
\end{equation}
where $N_{reco\ in\ i}^{true\ in\ j}$ is the number of events reconstructed in bin $i$ that had true kinematics 
corresponding to bin $j$, and $N^{true\ in\ j}$ is the number of events with true kinematics corresponding to bin $j$.
The equation for the 
predicted 
observed number of events, $N_{i}$,
in terms of the events in true kinematic bin $j$ becomes:

\begin{equation}
\begin{aligned}N_{i}= & \sum_{j}^{N_{bin}}S_{ij}\Bigg\{ c_{j}N_{j}^{sig,MC}\\
 & +\sum_{k}^{bkgd\ types}\left(\prod_{a}^{model}\omega\left(\vec{a}\right)_{j}^{k}\right)N_{j}^{bkgd\ k,MC} \Bigg\}
\end{aligned}
\label{eq:N_i}
\end{equation}

The above
Eq.\eqref{eq:N_i} 
forms a mapping between true bin $j$ and reconstructed
bin $i$. This approach\,\cite{Abe:2018dolan} after fitting the parameters,
will $unfold$ the true number of events $c_{j}N_{j}^{sig,MC}$ in
bin $j$ from the observed data. Using the histogram of observed reconstructed
events $N_{i}^{obs}$ and the predicted number of observed events $N_{i}(\vec{c},\vec{a})$
from Eq.\eqref{eq:N_i}, which depends on the fit parameters $c_{j}$
and model parameters $\vec{a}$, we can form the binned likelihood
of a histogram\,\cite{baker-cousins} as:
\begin{equation}
\begin{aligned}-2\ln(L)_{stat}= & \sum_{i}^{bins}2\bigg\{N_{i}(\vec{c},\vec{a})-N_{i}^{obs}\\
 & +N_{i}^{obs}\ln\left(\frac{N_{i}^{obs}}{N_{i}(\vec{c},\vec{a})}\right)\bigg\}
\end{aligned}
\label{eq:chisq-stat}
\end{equation}
which will be minimized.

In addition, three penalty terms are added to Eq.\eqref{eq:chisq-stat}.
The first is:
\begin{equation}
\begin{aligned}-2\ln(L)_{bkgd}= & \left(\vec{a}-\vec{a}{}_{prior}\right)^{T}\left[V_{cov}^{model}\right]^{-1}\\
 & \times\left(\vec{a}-\vec{a}_{prior}\right)
\end{aligned}
\label{eq:chi-bkgd}
\end{equation}
where $V_{cov}^{model}$ is a covariance matrix containing the uncertainties
and correlated errors on the background model parameters $\vec{a}$
and the initial parameter value is given as $\vec{a}_{prior}$ which has been
discussed in \cite{Abe:2017vif}. 


The number of observed events includes a flux term that is the number
of $\bar\nu_\mu$ per unit area. This has been modeled for the different
neutrino energies as:
\begin{equation}
\sum_{n}^{E_{\nu}}f_{n}^{i}
\end{equation}
where $f_{n}^{i}$ is the fraction of antineutrinos in flux energy bin $n$ for reconstructed
bin $i$. This nominally sums to unity. The flux uncertainty is given in a covariance
matrix $V_{cov}^{flux}$ and this adds to Eq.\eqref{eq:chisq-stat}
the flux penalty term: 

\begin{equation}
\begin{aligned}-2\ln(L)_{flux}= & \left(\vec{f}-\vec{f}{}_{prior}\right)^{T}\left[V_{cov}^{flux}\right]^{-1}\\
 & \times\left(\vec{f}-\vec{f}_{prior}\right)
\end{aligned}
\label{eq:chisq-flux-1}
\end{equation}

Finally the detector systematic uncertainties are given in a third
covariance matrix, $V_{cov}^{det}$, with $\vec{r}$ parameters which
vary the reconstructed event rate $r_i$ in bin $i$. This adds the last
penalty term given as:

\begin{equation}
\begin{aligned}-2\ln(L)_{det}= & \left(\vec{r}-\vec{r}{}_{prior}\right)^{T}\left[V_{cov}^{det}\right]^{-1}\\
 & \times\left(\vec{r}-\vec{r}_{prior}\right)
\end{aligned}
\label{eq:chisq-det}
\end{equation}

The measurement described here is concerned with events that occur specifically on water targets.  
The number of signal events occurring on water and non-water targets are allowed to vary independently 
in the fit so that the interaction rate on only water targets can be extracted.  We introduce a second set of 
scaling parameters, $d_j$ for events that occur on non-water targets: 

\begin{equation}
	\begin{aligned}N_{i}= &r_{i}\left(\sum_{n}^{E_{\nu}}f_{n}^{i}\right)\sum_{j}^{N_{bin}}S_{ij}\Bigg\{c_{j}N_{j}^{sig,water,MC}\\
		& +d_{j}N_{j}^{sig,non-water,MC}\\
 & +\sum_{k}^{bkgd\_types}\left(\prod_{a}^{model}\omega\left(\vec{a}\right)_{j}^{k}\right) N_{j}^{bkgd\_k,MC}\Bigg\}
\end{aligned}
\label{eq:N_i-prediction}
\end{equation}

Data samples where there was no water in the P$\emptyset$D bags serve to constrain the $d_j$ parameters so that
while simultaneously fitting water-in and water-out data, the unfolded CC-0$\pi$ event rate on water is extracted from the 
data as the $c_{j}N_{j}^{sig,water,MC}$ term.

The final log likelihood
equation of all terms that will be minimized to fit the data is:

\begin{equation}
\begin{aligned}-2\ln(L)_{tot}= & -2ln(L\left[\vec{c},\vec{d},\vec{a},\vec{f},\vec{r}\right])_{stat}\\
 & -2\ln(L\left[\vec{a}\right])_{bkgd}\\
 & -2\ln(L\left[\vec{f}\right])_{flux}\\
 & -2\ln(L(\left[\vec{r}\right]))_{det}
\end{aligned}
\label{eq:chisq-total}
\end{equation}
where the fit parameters dependence of each likelihood term is made
explicit. Note that ultimately, we are interested in the $\vec{c}$ fit parameters that will be used to extract the unfolded true differential
water cross section. This method differs from the D'Agostini iterative unfolding method used in \cite{Abe:2018tianlu}, which did a single iteration and did
not compare results with and without regularization.

\subsection{Binning Choice}

\begin{figure*}
\includegraphics[scale=0.42]{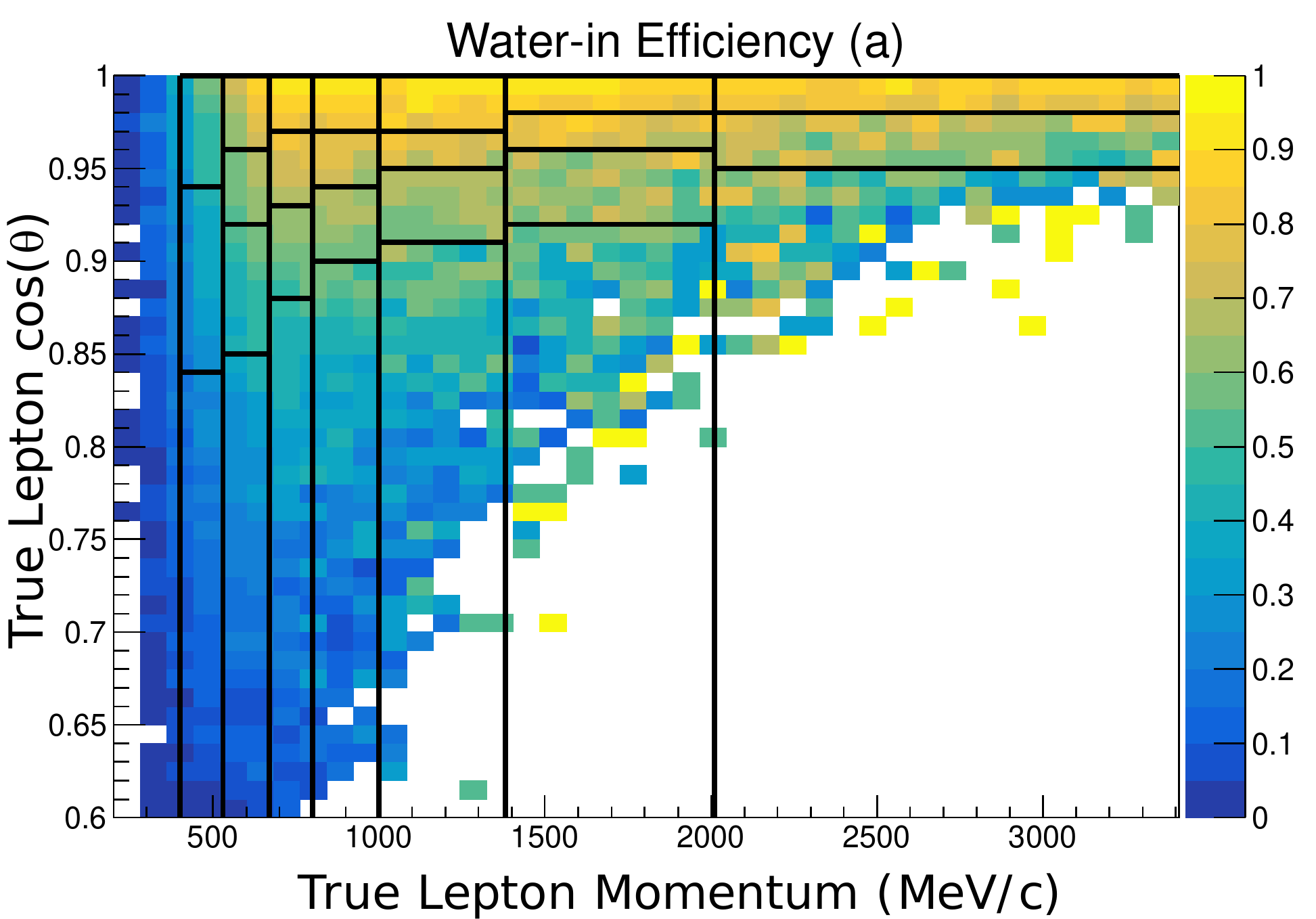}
\includegraphics[scale=0.42]{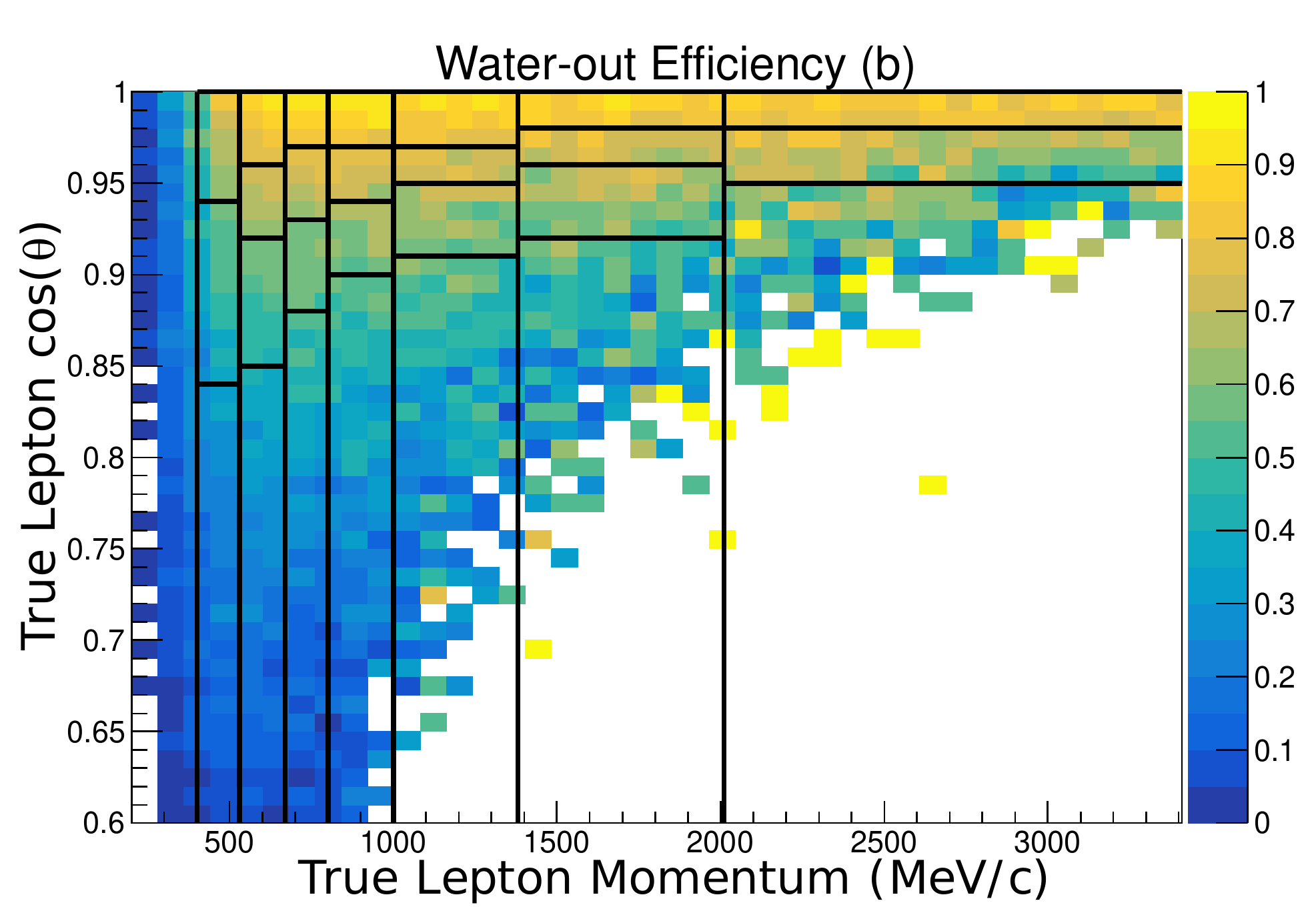}

\includegraphics[scale=0.42]{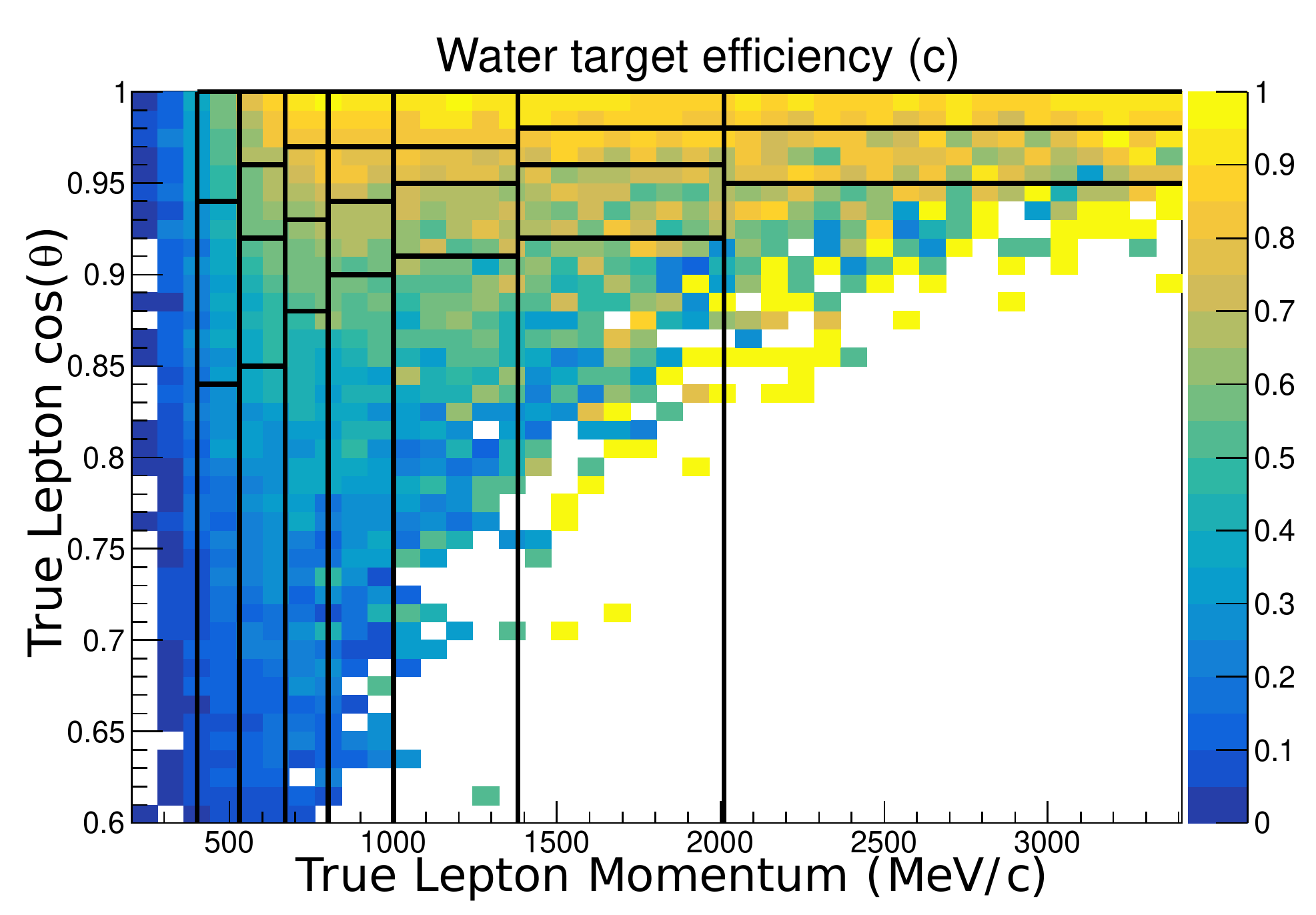}
\includegraphics[scale=0.42]{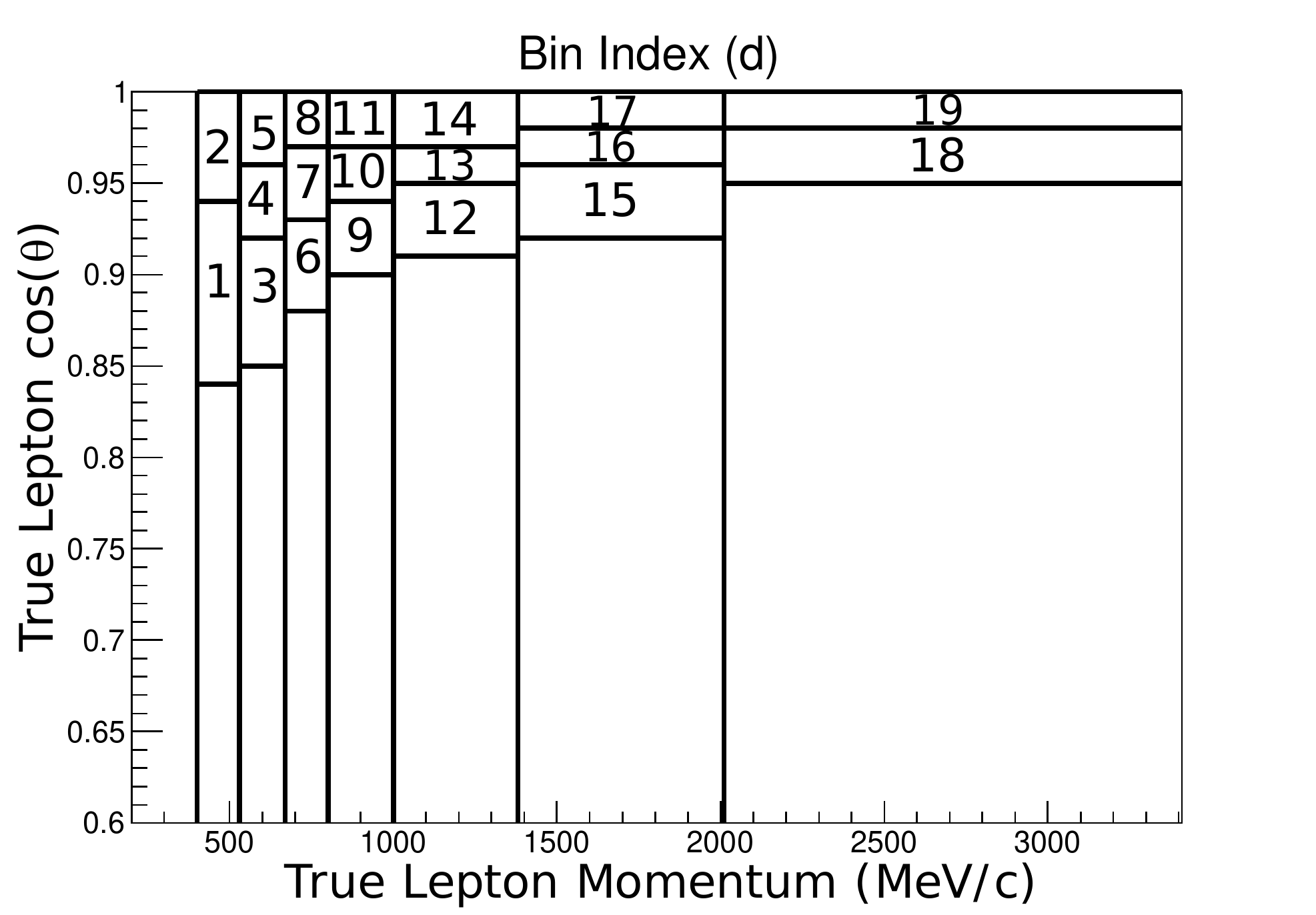}
\caption{\label{fig:Efficiency-plots}The CC-$0\pi$ selection efficiency plots in 2-D $p$ vs. $\cos\theta$ bins
for water-in (a), water-out (b) and water target only (c). There are 28 bins whose edges are drawn
with vertical and horizontal lines. The efficiencies are given in
color bands and it is noted that the efficiencies are very similar. 
The last plot (d) is the bin index given in Table IV. Note that the
28th bin in Table III is outside the plot boundary. 
The fit results in Section VI.A. use these 19 bins
which are a subset of the 28 bins.}
\end{figure*}

The choice of the 2 dimensional $\mu^+$ track $p$-$\cos\theta$ binning
was determined by the following considerations:
\begin{enumerate}
\item The number of events in each 2-D bin should have reasonable statistics,
$\sim$100 events. This improves the stability of the 
fit results. 
\item The selection efficiency should be relatively high to minimize model dependence of the 
	efficiency correction, and event populations
should not differ very much between adjacent bins which also improves the stability of the 
fit results.
\item The bin sizes should be fine enough so local detector resolution effects
are well represented and the detector resolutions do not
change too much from bin to bin, however not too fine such that
there are too few events in the bin. 
\end{enumerate}
We expect these choices should reduce regularization complications,
which are discussed in later sections, or possibly even the need for
regularization. The 28 bins over the entire kinematic phase space
are specified in Table \ref{tab:Binning}. The 2-D plot in Fig.
\ref{fig:Efficiency-plots} contains the efficiencies of the water-in
(a) and water-out (b) data sets. 
\begin{table}
\begin{tabular}{ccc}
\hline 
Bin  & True Momentum & True $\cos\theta$\tabularnewline
Index & MeV/$c$ & Bin edge\tabularnewline
\hline 
\hline 
1 & 0-400 & -1,1\tabularnewline
2-4 & 400-530 & -1,0.84,.94,1\tabularnewline
5-8 & 530-670 & -1,0.85,0.92,0.96,1\tabularnewline
9-12 & 670-800 & -1,0.88,0.93,0.97,1\tabularnewline
13-16 & 800-1000 & -1,0.90,0.94,0.97,1\tabularnewline
17-20 & 1000-1380 & -1,0.91,0.95,0.97,1\tabularnewline
21-24 & 1380-2010 & -1,0.92,0.96,0.98,1\tabularnewline
25-27 & 2010-3410 & -1,0.95,.98,1\tabularnewline
28 & 3410-50000 & -1,1\tabularnewline
\hline 
\end{tabular}

\caption{\label{tab:Binning} $p$-$\cos\theta$ bins over all kinematic phase
space}
\end{table}

Among the 28 bins covering the entire kinematic region, there are bins
that have very few events due to the phase space or due to the low
detector efficiency. These include the first ($p<400$ MeV/$c$) and
last ($p>3410$ MeV/$c$) bins and lowest lying $\cos\theta$ bins in
each of the seven given momentum slices in the middle momentum ($400<p<3410$
MeV/$c$) bins. Although we will fit in all 28 bins, we do not use these
nine bins in the final differential cross section determinations.
Instead we use the other 19 bins for the final differential cross
section measurements. These 19 cross section bins are given in Table
\ref{tab:Analysis-bins} and their index number is called a cross section bin.

\begin{table}
\begin{tabular}{ccc}
\hline 
Bin  & Momentum & $\cos\theta$\tabularnewline
Index & MeV/$c$ & Bin edge\tabularnewline
\hline 
\hline 
1,2 & 400-530 & .84,.94,1\tabularnewline
3,4,5 & 530-670 & .85,.92,.96,1\tabularnewline
6,7,8 & 670-800 & .88,.93,.97,1\tabularnewline
9,10,11 & 800-1000 & .90,.94,.97,1\tabularnewline
12,13,14 & 1000-1380 & .91,.95,.97,1\tabularnewline
15,16,17 & 1380-2010 & .92,.96,.98,1\tabularnewline
18,19 & 2010-3410 & .95,.98,1\tabularnewline
\hline 
\end{tabular}

\caption{\label{tab:Analysis-bins}$p$-$\cos\theta$ bins used for the unfolded
cross sections and indexed as cross section bin numbers.}
\end{table}

\subsection{Fit Parameters, Systematic Errors, and Checks}

The parameters used in the likelihood fit 
in Eq.\eqref{eq:chisq-total}\\
include the signal interactions on water targets coefficients $\overrightarrow{c}$, the 
signal interactions on non-water targets
coefficients $\overrightarrow{d}$, the fractional flux parameters
$\overrightarrow{f}$, the background model parameters $\overrightarrow{a}$,
and the reconstructed event rate scale factor $\overrightarrow{r}$. 
All types of parameters are listed with their numbers
in Table \ref{tab:Parameters}. We describe each parameter type in
the following paragraphs.

There are 
two sets of 
28 scale factors for the $p-\cos\theta$ bins, one set $\overrightarrow{c}$
for interaction on water and another $\overrightarrow{d}$ for non-water interactions. The
water parameters, $\overrightarrow{c}$, contain the subset of 19
parameters that are used to extract the final unfolded cross section.

There are 11 flux parameters representing the fraction of the $\overline{\nu}_{\mu}$
flux in varying energy bin widths with energy boundaries at 0, 0.4,
0.5, 0.6, 0.7, 1.0, 1.5, 2.5, 3.5, 5.0, 7.0, and 30.0 GeV. The pre-fit
flux uncertainties are on the order of $\sim10\%$ in the matrix $V_{cov}^{flux}$.

There are 9 background model parameters and 6 pion final state interaction (FSI) parameters.
The first three background model parameters, the axial mass, the axial form factor, and the fraction of non-resonant background, describe the main background, which is the charged current resonant background. 
The charged current deep inelastic background is described using a scaling parameter on a normalization 
function of the cross-section, which depends on the neutrino energy.
The other background model parameters are normalization rates for the charged current coherent interactions on Carbon and Oxygen, neutral current, and coherent neutral current backgrounds.
More details about those parameters can be found in \cite{Abe:2017bay}.

The 6 pion FSI parameters include effects
for absorption, inelastic scattering, charge exchange, and quasielastic
scattering inside the nucleus. For descriptions of these FSI parameters see Table IV
in a previous T2K publication \cite{Abe:2015oar}. 

The detector 
parameters $\overrightarrow{r}$ scale the predicted
number of reconstructed events in Eq.\eqref{eq:N_i-prediction} in
each bin $i$ of reconstructed $\mu^+$ kinematics. These parameters also
are included in the penalty terms in Eq.\eqref{eq:chisq-det} and,
being scale factors, they are nominally set
to 1.0. There is one parameter for each of the
19 cross section bins for each water-in/water-out samples of the CC-0$\pi$
and CC-1$\pi$ selections. 
This totals to 76 detector parameters.
The uncertainties of these parameters are determined from detector
uncertainties in the TPC and the P$\emptyset$D detectors. The TPC
and P$\emptyset$D momentum resolution and scale errors and the B-field
distortions are estimated by varying their scales resulting in their
combined errors of roughly 6\%. The TPC charge mis-identification, track
reconstruction efficiency, shower reconstruction efficiency, and TPC-P$\emptyset$D
matching errors are obtained by reweighting the parameters, resulting
in their combined error of roughly 2.5\%. 
The efficiency dependence on the signal CC-$0\pi$ model parameters was checked
by varying the CCQE axial mass and Carbon and Oxygen antineutrino interaction signal model parameters. 
The remaining errors are due to
the uncertainty on the mass of the non-water material in the P$\emptyset$D
detector \cite{Abe:2018tianlu} which was estimated to be $1.5\%$ and the
mass of water of the filled water target bags. The uncertainty of
the water mass in each P$\emptyset$D water-bag was modeled by an uncorrelated 
normal distribution with a 10\% standard deviation. The typical initial errors
on the parameters representing the CC-0$\pi$ samples are 5-10\% whereas
the errors on the CC-1$\pi$ samples are 10-20\%.

\begin{table}
\begin{tabular}{ccc}
\hline 
Symbol & Parameter & Number\tabularnewline
\hline 
\hline 
	$\overrightarrow{c}$ & signal on water coefficients & 28\tabularnewline
	$\overrightarrow{d}$ & signal on non-water coefficients & 28\tabularnewline
$\overrightarrow{f}$ & flux parameters & 11\tabularnewline
$\overrightarrow{r}$ & detector parameters & 76\tabularnewline
$\overrightarrow{a}$ & background and FSI parameters & 15\tabularnewline
\hline 
\end{tabular}

\caption{\label{tab:Parameters}Table of parameters in the fit.}
\end{table}

Basic validation checks, that the fit behaves properly under the 
conditions that the MC matches the data with well defined conditions, were performed.
The first check
consisted of fitting the NEUT MC model to verify that all the fitted
water coefficients, $c_j$, and non-water coefficients, $d_j$,
are exactly reproduced. The next check was to decrease/increase the
water/non-water target masses by $\pm$50\% and check that the
$c_j$ and $d_j$ parameters decrease/increase by the correct
amount. 

The systematic errors on the flux, background parameters and detector
systematics, which appear in the penalty terms in Eqs.
(\ref{eq:chi-bkgd}), (\ref{eq:chisq-flux-1}) and (\ref{eq:chisq-det}),
were checked by removing 2 of the 3 groups of nuisance parameters and checking
the values of the refit water-in coefficients. When each of these groups
are turned on and off one by one, we find that water-in coefficients
have errors in the range of 2-6\%, 2-6\%, and 6-14\% due to uncertainties on the
flux, background models, and detector systematics, respectively.

Finally, five different samples of the NEUT MC model, with the same
number of events as the expected data sample, were generated and fitted.
The resulting water coefficients $c_j$ were all consistent between
all five samples. 
To evaluate how well the post-fit results agree with a certain prediction, we define the $\chi^2$
between some prediction
with label A and the post-fit results to be:
\begin{equation}
	\label{eq:chi2ofFit}
	\chi^2_{A} = (\vec{\sigma}_{A} - \vec{\sigma}_{\text{post-fit}})^T\left[V_{cov}^{\text{post-fit}}\right]^{-1}(\vec{\sigma}_{A} - \vec{\sigma}_{\text{post-fit}})
\end{equation}

The resulting $\chi^2$s between the MC true event rates and the fitted ones from the five different samples
had similar values. 

\subsection{Regularization}

\begin{figure}
\includegraphics[scale=0.4]{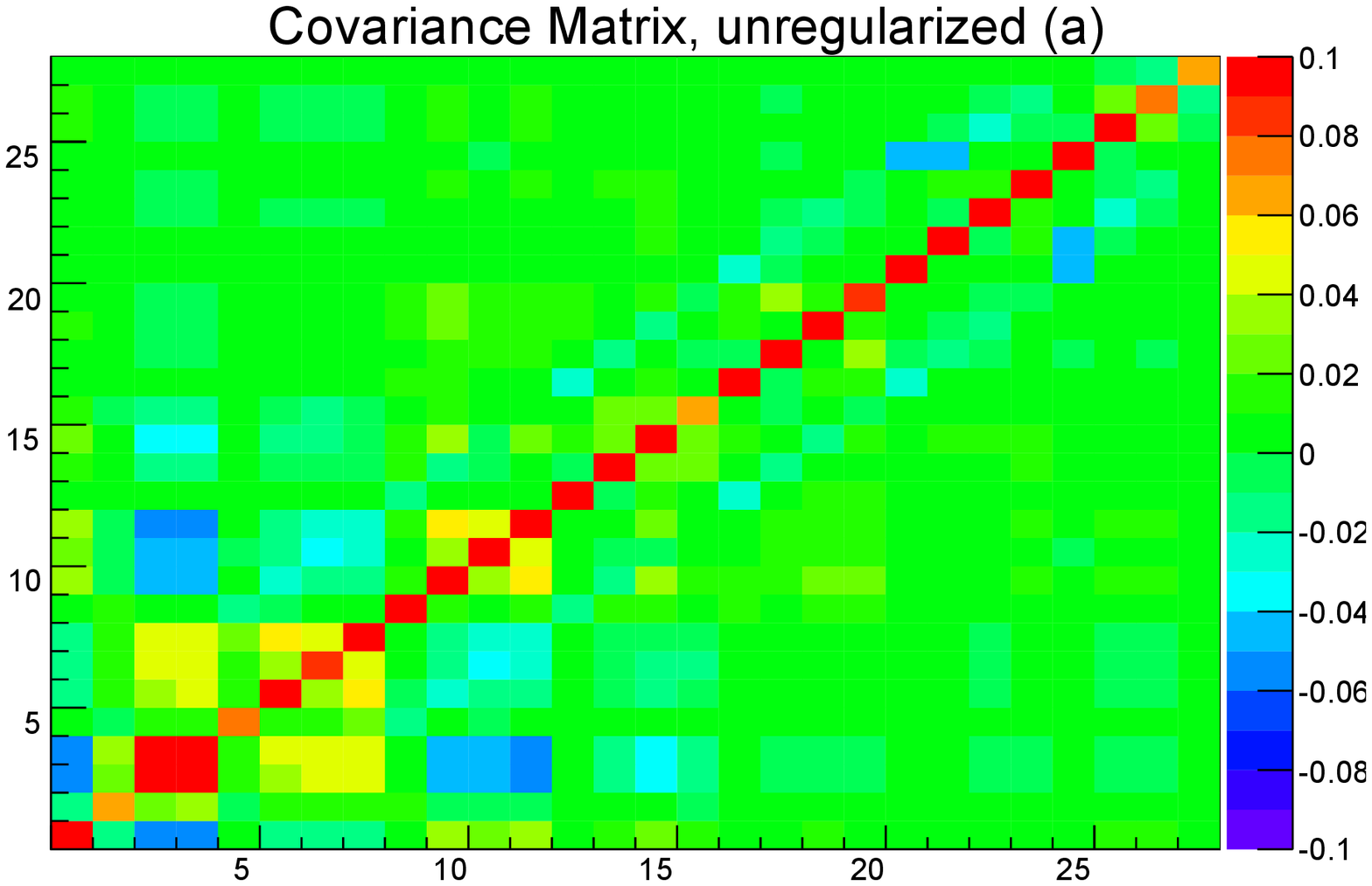}

\includegraphics[scale=0.4]{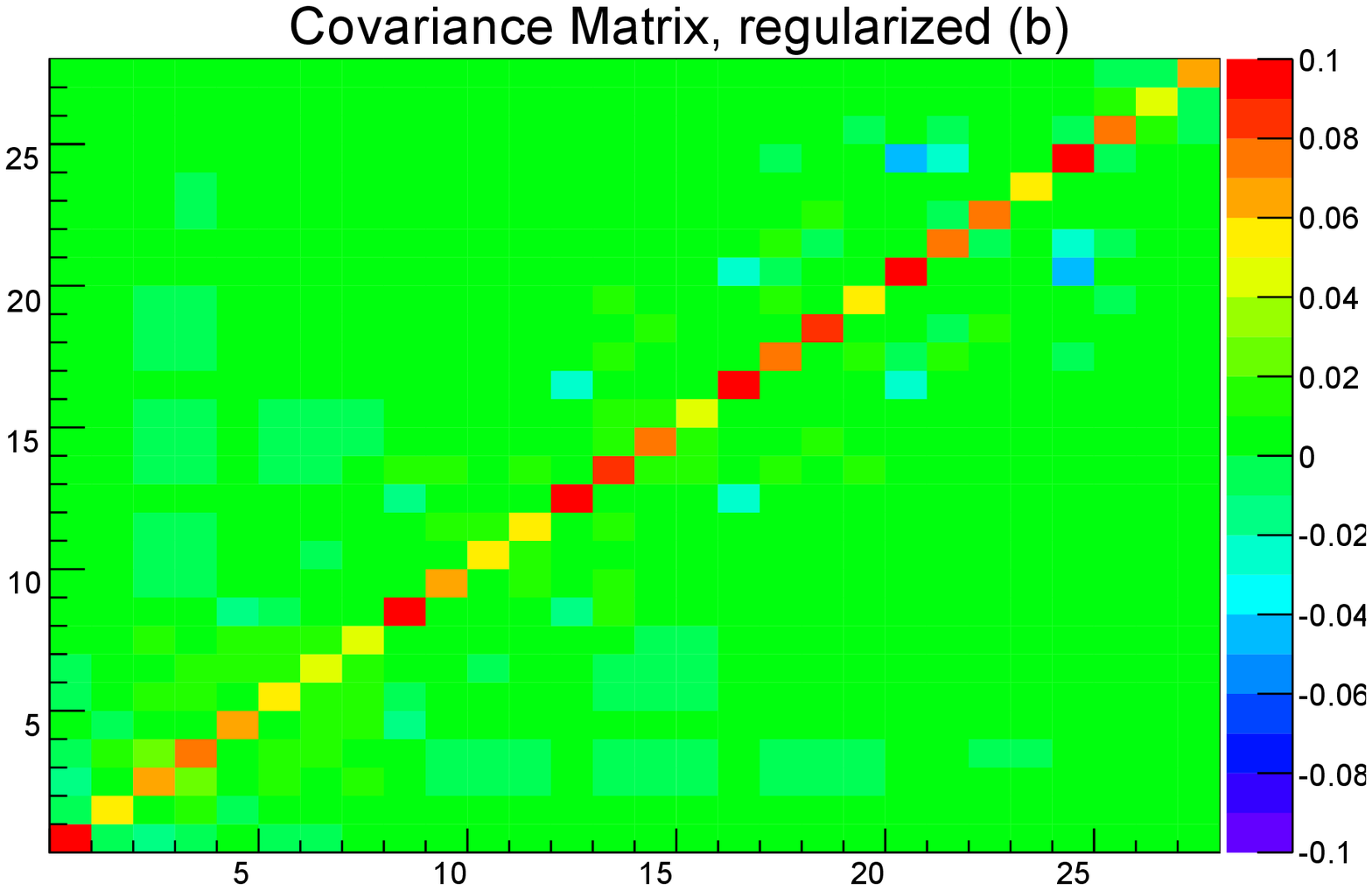}

\caption{\label{fig:fit-param-cov}Covariance Matrix of water-in coefficients
before (a) and after (b) regularization was applied to a test MC
sample. The regularization reduces off-diagonal correlations.}
\end{figure}

The aim of the analysis is to extract the parameters $c_j$ which
are proportional to the number of CC-0$\pi$ events on water in the $p-\cos\theta$ bins
for i=1,...,28. This is obtained by fitting the parameters $c_{j}$ in Eq.\eqref{eq:N_i-prediction} 
which determines the predicted $N_{i}$
that is used in the binned likelihood in Eq.\eqref{eq:chisq-stat}
and Eq.\eqref{eq:chisq-total}. This forms an inverse problem where small
statistical fluctuations in the reconstructed event rates, $N_{i}$, can cause large variations of the fitted
parameters $c_{j}$. The Fig.\ref{fig:fit-param-cov}(a) shows the
covariance matrix of the fitted parameters $c_{j}$ using a MC simulation
test sample. There are some moderate bin to bin correlations seen
in this covariance matrix.  Specifically, there are off-diagonal anti-correlations between neighboring momentum bins for 
equivalent $\cos\theta$ bins.  These are caused by the fit being able to adjust the event rates in neighboring true bins in an 
anti-correlated way and getting similar predictions in the reconstructed bins.

These bin to bin variations can be reduced by applying data-driven
regularization methods as discussed and applied in Section IV.D in
a previous T2K analysis \cite{Abe:2018dolan}. 
The regularization technique \cite{calvetti}, 
consists of adding to Eq.\eqref{eq:chisq-total} an additional penalty term:
\begin{equation}
	-2log(L\left[\vec{c},p_{reg}\right])_{reg}=p_{reg}\sum_{i}^{N_{bin}-1}\left(c_{i}-c_{\hat{i}}\right)^{2}\label{eq: chi-regularization}
\end{equation}
where $\hat{i}$  
is the index of bin corresponding to a neighboring momentum bin $i$ for equivalent $\cos\theta$ bins.
Eq.\eqref{eq: chi-regularization} includes a parameter $p_{reg}$ that controls the regularization
strength between momentum bin boundaries. 
When Eq.\eqref{eq: chi-regularization} is added to Eq.\eqref{eq:chisq-total}
and the sum is minimized, this will clearly reduce variations between
adjacent momentum bins depending on the size of $p_{reg}$. The L-curve regularization 
\cite{lcurve} is obtained 
when the ratio 
$-2log(L\left[\vec{c},p_{reg}\right])_{reg} / p_{reg}$
has the largest curvature as a function of $p_{reg}$\cite{lcurve}.  
The $p_{reg}$ values of $1-2$ were found to have the largest curvature in this
test sample shown in Fig.\ref{fig:fit-param-cov}(a). When regularization
with $p_{reg}=1$ is applied to the test sample, the off-diagonal
covariances and the bin to bin correlations are reduced as shown in
Fig.\ref{fig:fit-param-cov}(b). 

Both unregularized and regularized results will be shown. 
They are expected to be totally equivalent in terms of physics
results but regularized results will 
minimize  unphysical large bin-to-bin fluctuations.
The purpose here
is to provide at the same time fully correct and model independent
results (unregularized) 
which are properly interpreted together
with a full covariance matrix provided in a data release.

\section{Data Results and Comparison to Models}

\subsection{Fit Results}

\begin{figure}
\includegraphics[scale=0.4]{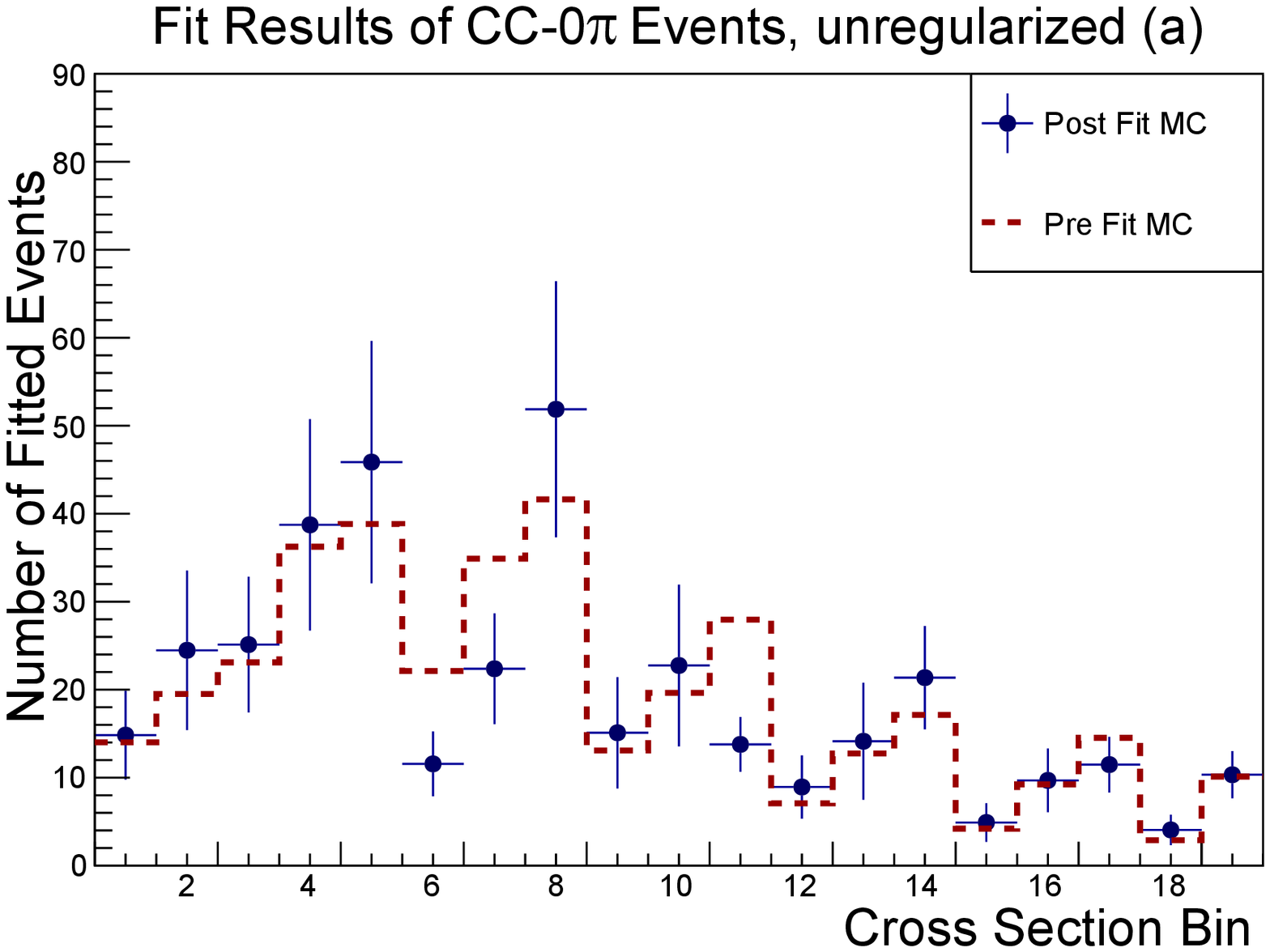}
\includegraphics[scale=0.4]{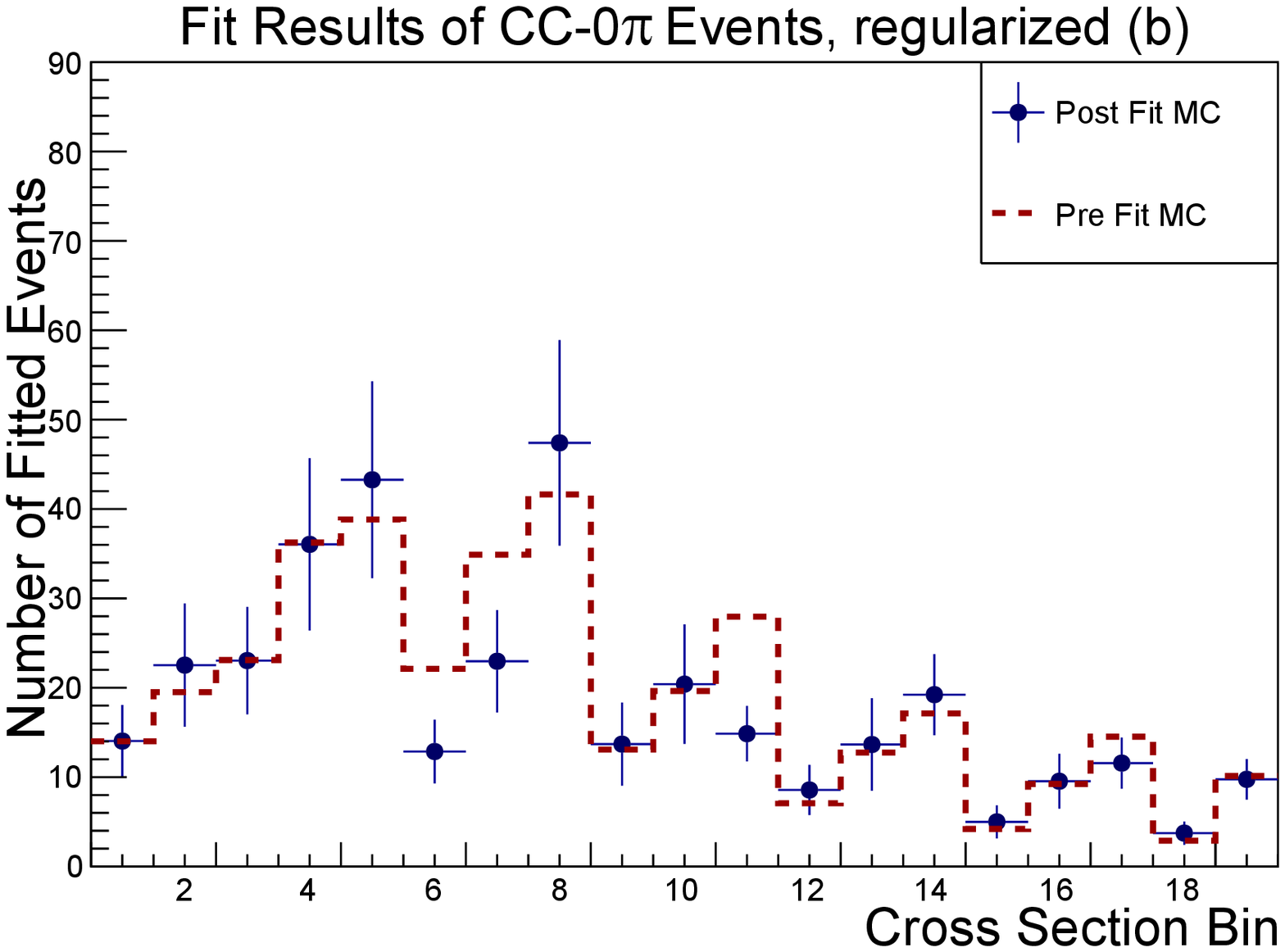}
\includegraphics[scale=0.4]{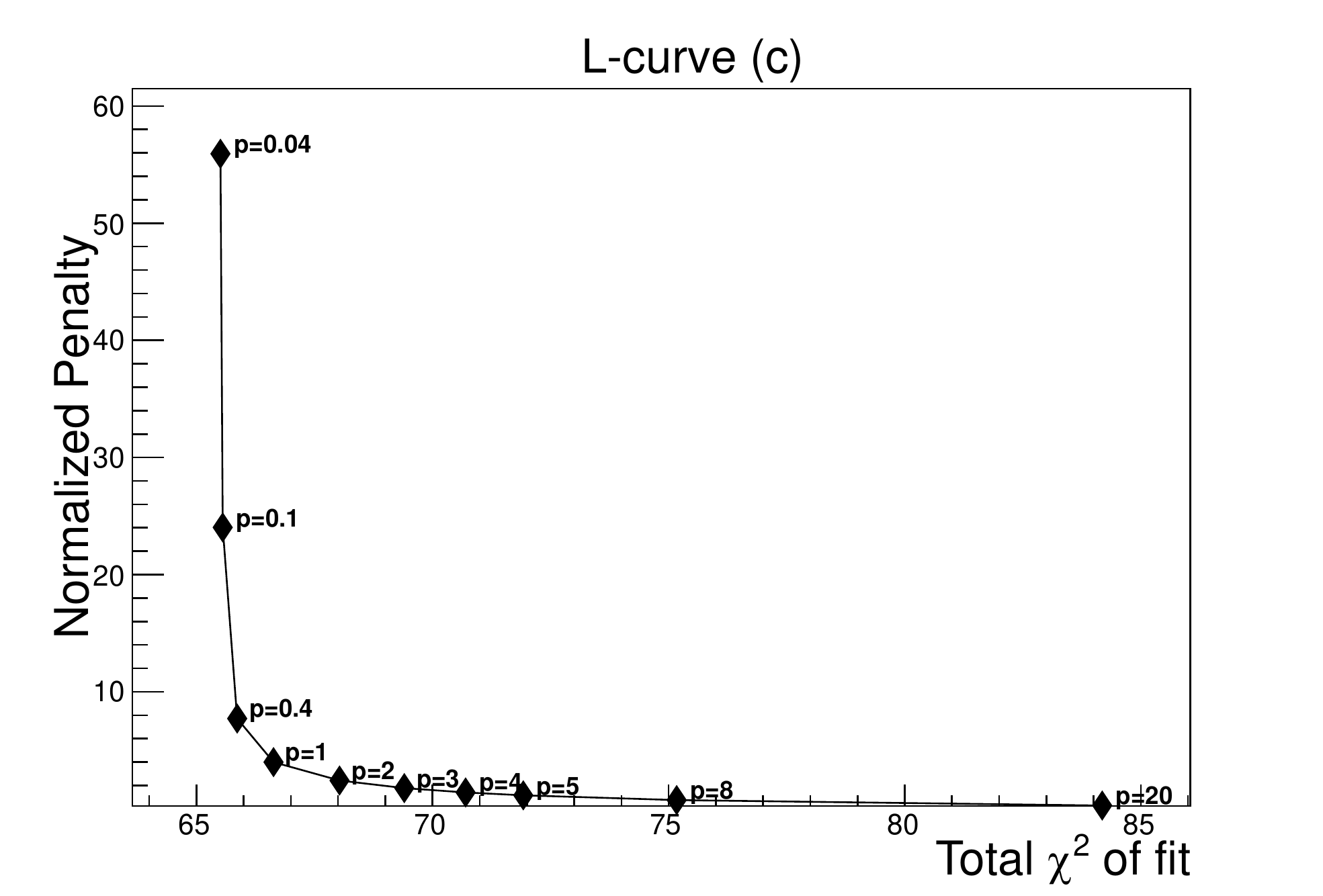}

\caption{\label{fig:fit_water-in_rates_nonreg+reg}Fit results of CC-$0\pi$ events rates
in 19 cross section bins for unregularized (a) and regularized (b) 
for water events and the regularization L-curve of data (c).} 
\end{figure}

The unregularized and regularized fit results of event rates with
errors for the 19 bins of the water 
CC-0$\pi$ cross section by cross section bin number
are shown in Fig. \ref{fig:fit_water-in_rates_nonreg+reg} (a)
and (b), respectively. The L-curve of the regularized fits is
shown in Fig. \ref{fig:fit_water-in_rates_nonreg+reg} (c). The
largest L-curvature occurs in data at $1$, and we choose $p_{reg}=1$
for the regularization. 

\begin{figure*}
\includegraphics[scale=0.4]{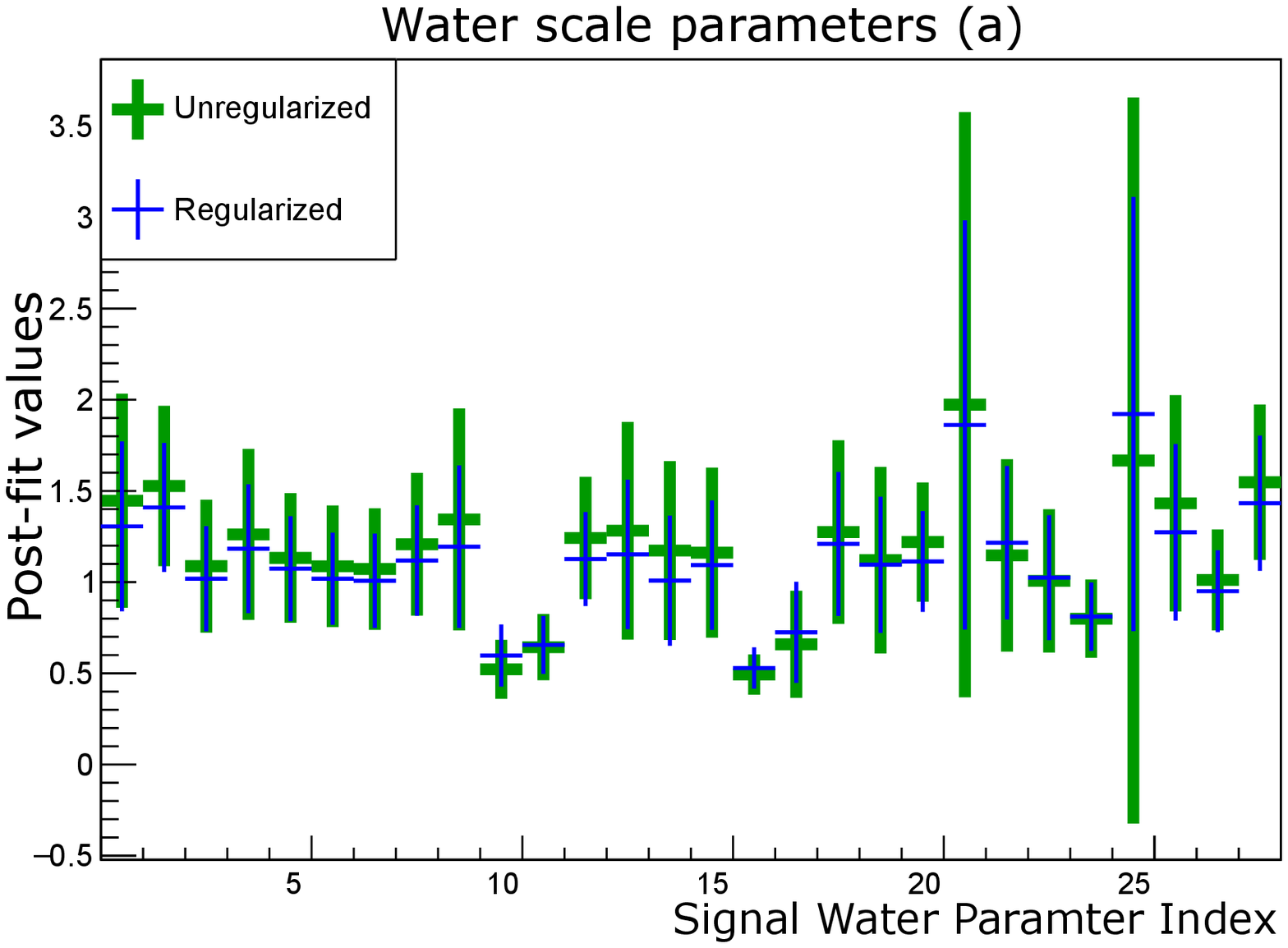}
\includegraphics[scale=0.4]{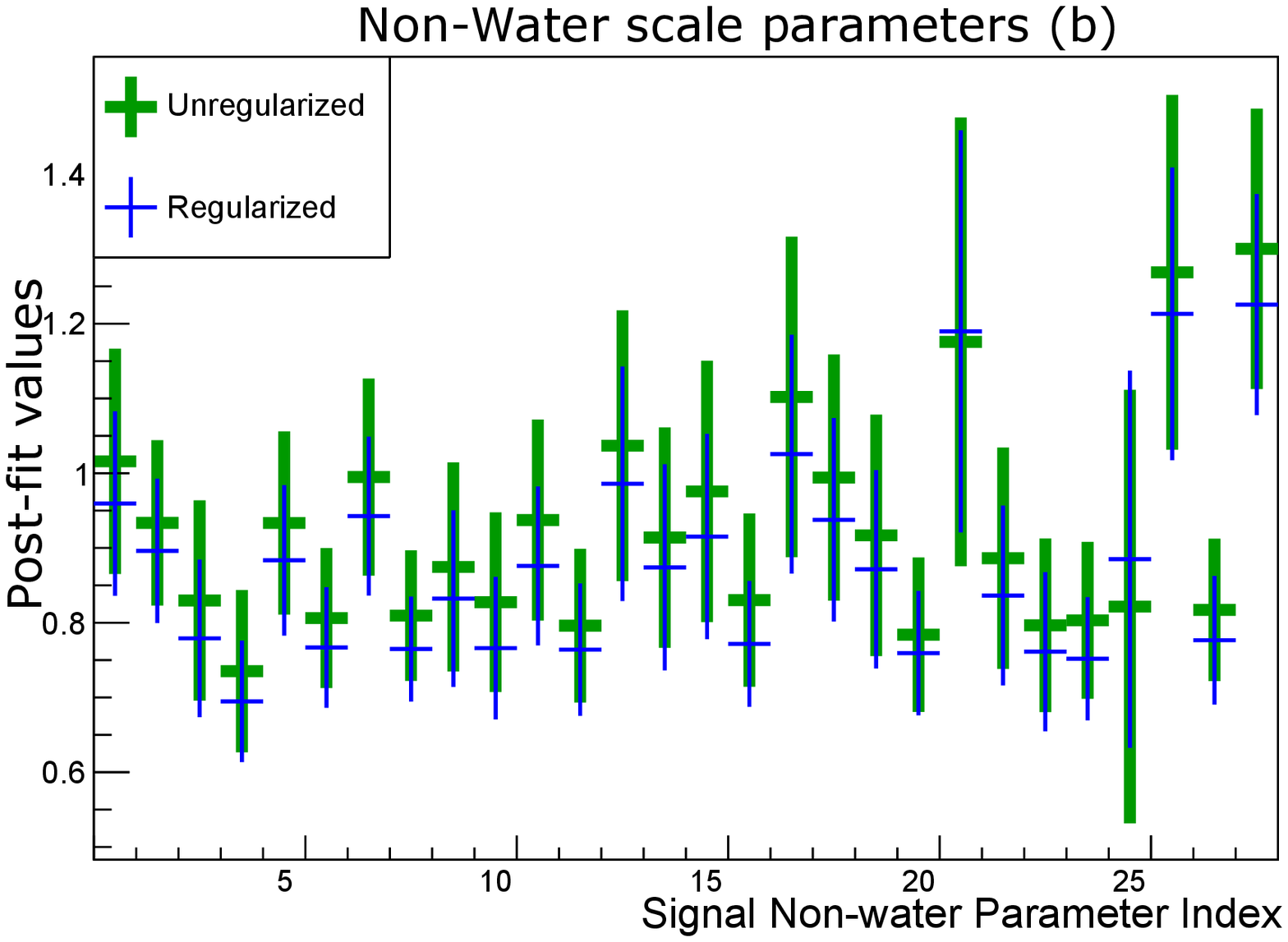}
\caption{\label{fig:Postfit_water-in/out_nonreg+reg}Post-fit results of 
water (a) and 
non-water (b) events 
which correspond to the 28 scale parameters $c_j$ and $d_j$, respectively
}
\end{figure*}

The resulting fitted or post-fit results for the 28 water 
$c_{j}$ and 28
non-water $d_{j}$ parameters are shown in Fig. 
\ref{fig:Postfit_water-in/out_nonreg+reg}
(a) and (b) respectively. The unregularized fit is in green
and the regularized fit is in blue. The nominal initial values are
set to 1.0, so the shifts or deviations from initial to post-fit values
can be readily inspected. The 
post-fit $c_{j}$ are centered on
$\sim1$ except for three (6th, 7th, and 11th) bins.
We note the 
non-water $d_{j}$ parameters are centered $\sim0.9$,
however, those same 3  bins
in the post-fit non-water parameters do
not have dips relative to their adjacent bins.

\begin{figure*}
\includegraphics[scale=0.4]{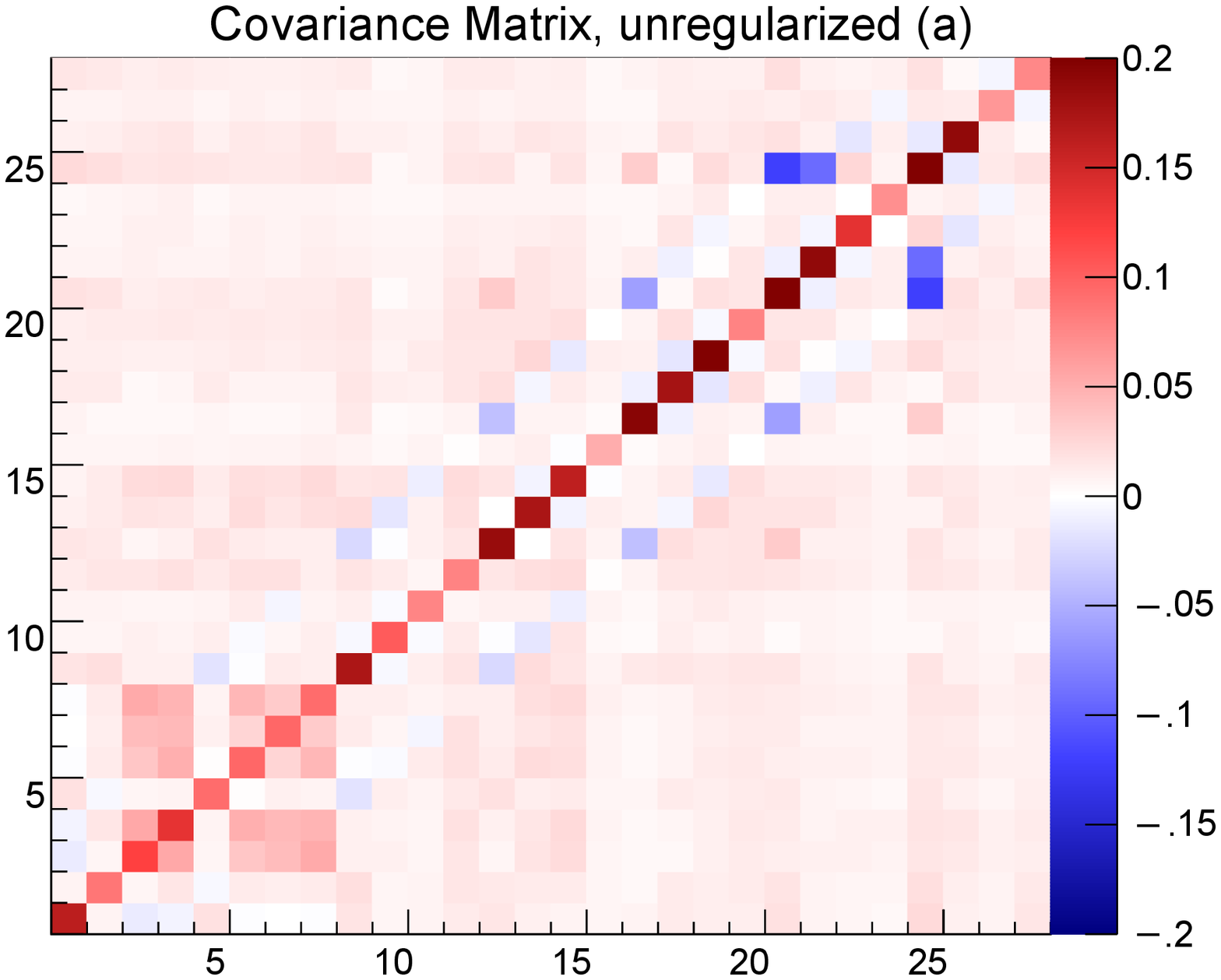}
\includegraphics[scale=0.4]{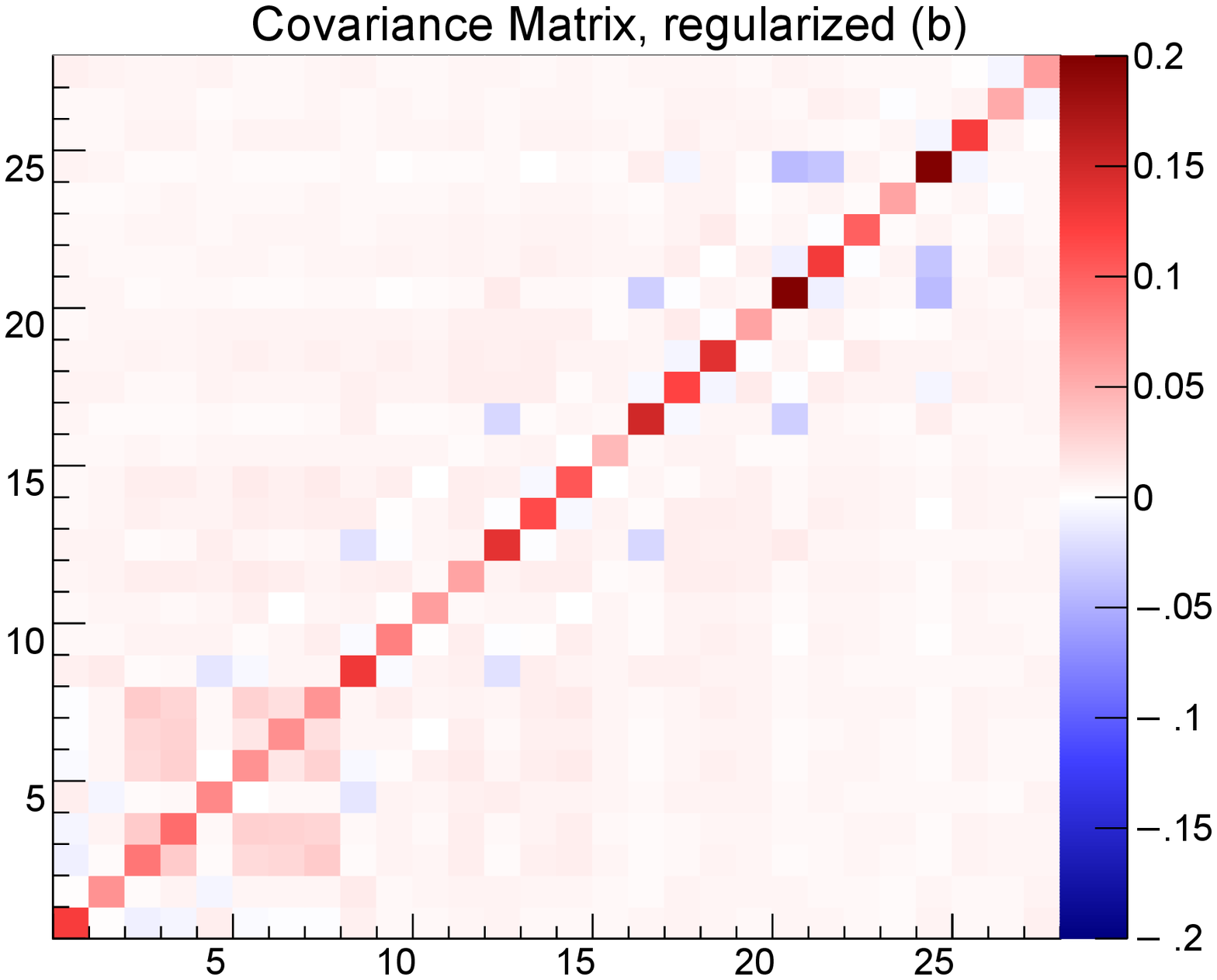}

\caption{\label{fig:CovMat_unreg+reg}Covariance Matrix of water parameters
for unregularized fits (a) and regularized fits (b).
} 
\end{figure*}

The covariance matrix of the fit results of the water $c_{j}$
parameters are shown in Figs. \ref{fig:CovMat_unreg+reg} 
(a) and (b) for unregularized and regularized fits, respectively. 
We observe in the unregularized
covariance slight positive (red bins) covariance correlations at low
momentum ($p<.67$ GeV/$c$) and a negative (blue bins) correlation in
bin 25 which is a high momentum ($p>2.01$ GeV/$c$) bin.

\subsection{Cross Section Comparisons to NEUT and other Models }

\begin{figure*}
\includegraphics[scale=0.8]{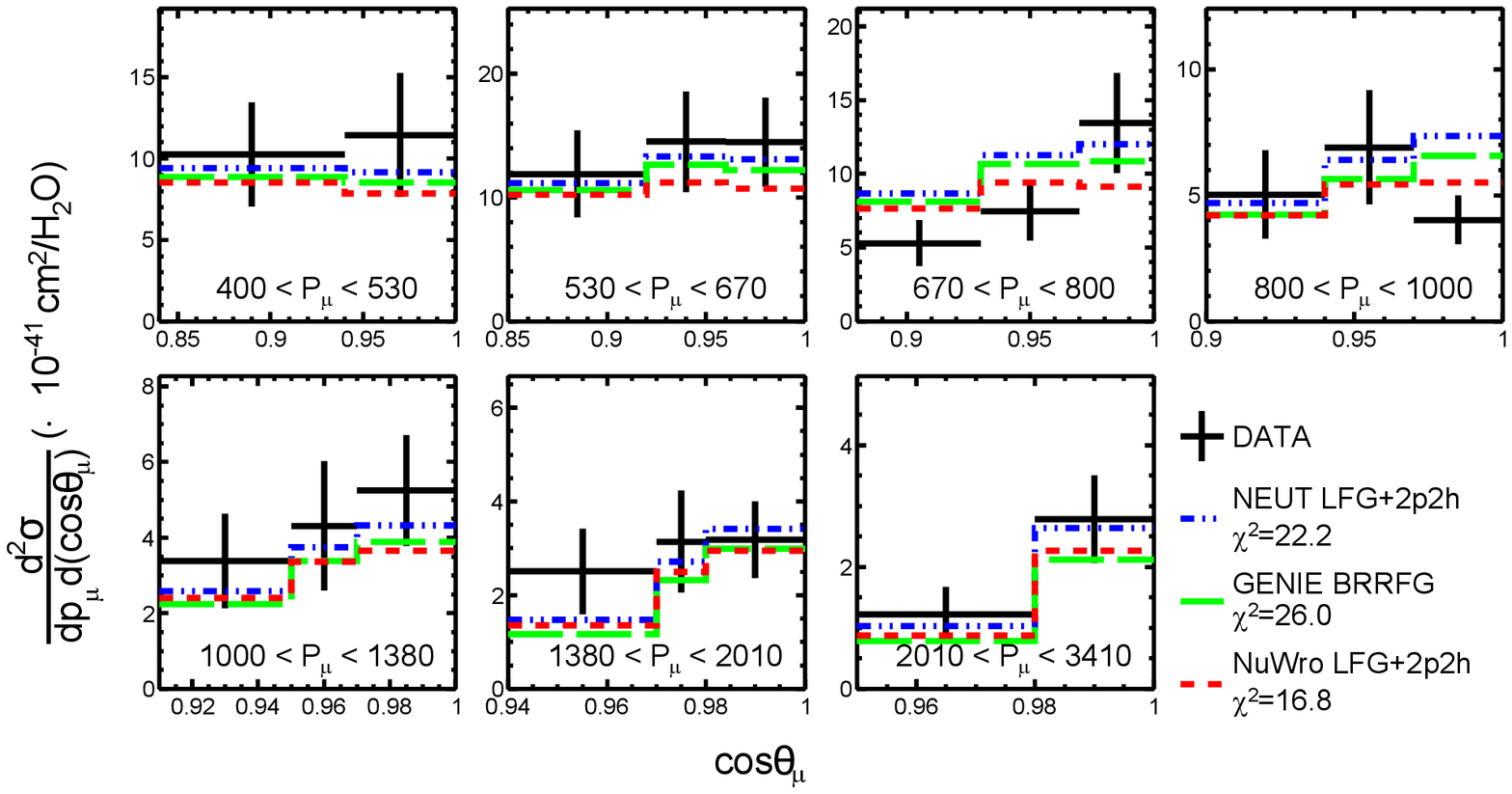}

\caption{\label{fig:Regularized-fit-results} Regularized fit
results of data as a function of 19 $\cos\theta$ bins in seven different
momentum ranges with comparisons to 
NEUT(ver.5.41), GENIE(ver.2.12.10), and NuWro(ver.18.02.1) predictions.
The fit $\chi^2$'s of each model is defined by Eqn. 13.}
\end{figure*}

\begin{figure*}
\includegraphics[scale=0.8]{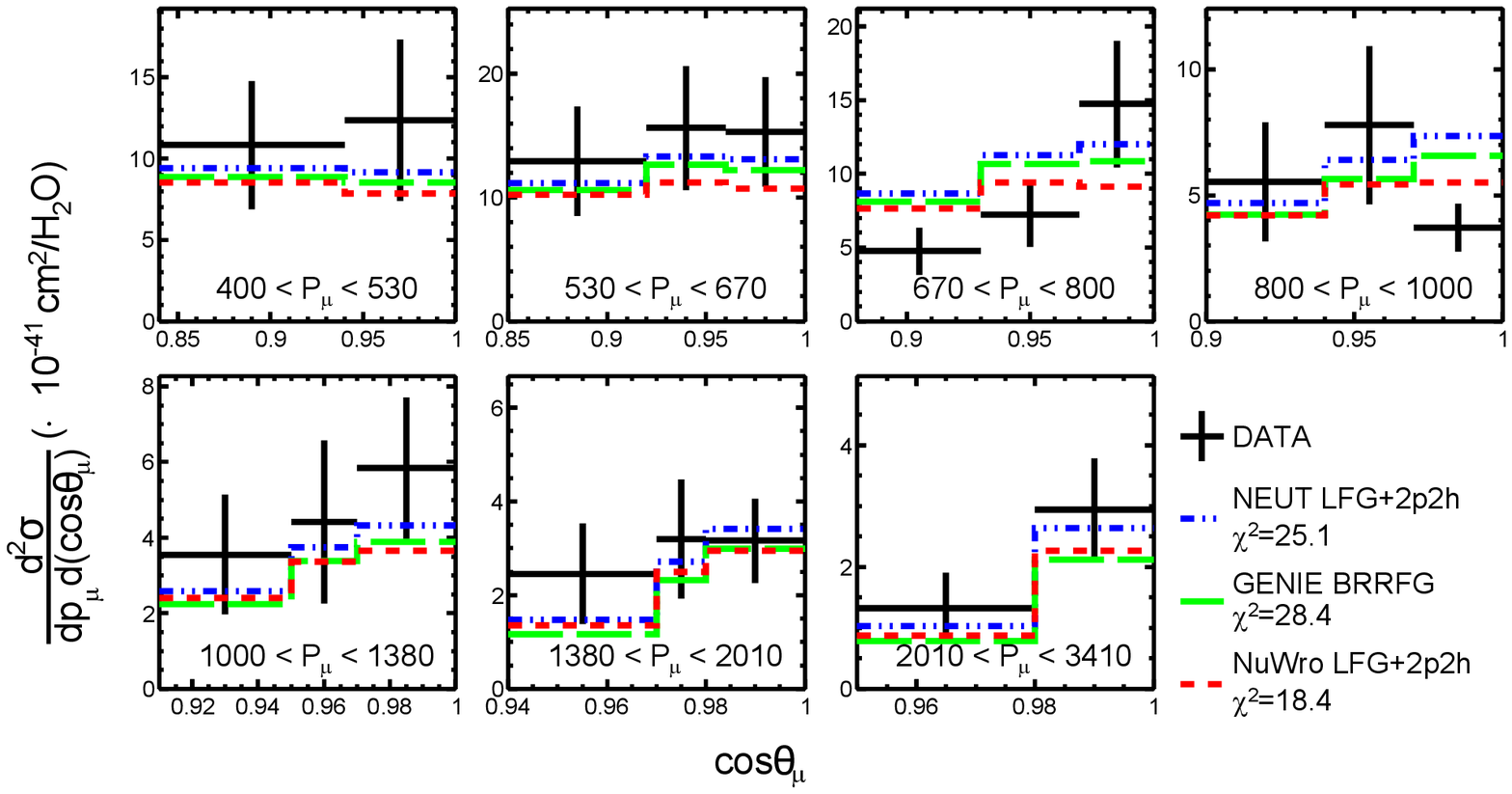}

\caption{\label{fig:Fit-results} Unregularized fit results
on data as a function of 19 $\cos\theta$ bins in seven different
momentum ranges with comparisons to 
NEUT(ver.5.41), GENIE(ver.2.12.10), and NuWro(ver.18.02.1) predictions.
The fit $\chi^2$'s of each model is defined by Eqn. 13.}
\end{figure*}

The regularized and the unregularized fit results of unfolded $p$
vs $\cos\text{\ensuremath{\theta}}$ bins of data (black crosses) with
comparisons to cross section predictions from
NEUT (ver.5.41), GENIE (ver.2.12.10), and NuWro (ver.18.02.1) models are shown in Fig. \ref{fig:Regularized-fit-results}
and Fig. \ref{fig:Fit-results}, respectively. 

The NEUT and NuWro models both include Local Fermi Gas (LFG) with 2p2h and the
GENIE model includes the Bodeck-Ritchie modifications to the Relativistic Fermi gas effects. These models have been
described in a previous T2K publication\cite{Abe:2018dolan} and the models 
were implemented using the NUISANCE framework\cite{Stowell:2016jfr}.
The results are presented
in seven plots of $\cos\theta$ bins in seven different momentum ranges
from 0.4 GeV/$c$ to 3.41 GeV/$c$. 
The data mostly agrees within 1 standard
deviation of all three predictions except for the 6th, 7th,
and 11th data bins that are $\sim2$ standard deviations below the NEUT prediction.
These correspond to the 3 low bins 
numbers 6,7, and 11 in Fig. \ref{fig:fit_water-in_rates_nonreg+reg},
numbers 10,11, and 16 in Fig. \ref{fig:Postfit_water-in/out_nonreg+reg},
and the $670<p<800$ MeV/$c$ (1st and 2nd bin) and $800<p<1000$ MeV/$c$ (3rd bin) in
Figs \ref{fig:Regularized-fit-results}  and \ref{fig:Fit-results}. 

\begin{table}
\begin{tabular}{ccc}
\hline 
	Generator & data $\chi^2$ & data $\chi^2$ \tabularnewline
	& (regularized) & (unregularized)\tabularnewline
\hline 
\hline 
	NEUT & 29.2 & 33.1\tabularnewline
	GENIE & 26.0 & 28.4\tabularnewline
	NuWro & 16.8 & 18.4\tabularnewline
\hline 
\end{tabular}

\caption{\label{tab:chi2toData}Comparisons of the data result in both the regularized and unregularized cases
	to NEUT, GENIE, and NuWro using the absolute $\chi^2$ from Eq.\eqref{eq:chi2ofFit}.}
\end{table}
The number of differential cross section bins, 19, is
the number of degrees of freedom in the $\chi^2$ comparisons in Table \ref{tab:chi2toData}.
We see generally good agreement with all three models, but a slight preference for the NuWro
prediction that has a lower $\chi^2=18.4$ for 19 degrees of freedom.  
In addition, the $\chi^2$'s between the regularized and unregularized cases are seen to be consistent.

The total cross section integrated over all 19 bins, can be determined
from the data and compared to NEUT, GENIE, and NuWro predictions. The T2K flux averaged
cross sections, in the kinematic phase space in Table IV, are given in units of 
$10^{-38}\frac{cm^{2}}{\mathrm{water\ molecule}}$ as, 


\begin{equation}
	\begin{split}
\sigma_{DATA}^{regularized} & = 1.11\pm0.18\\
\sigma_{DATA}^{unregularized} & = 1.17\pm0.22\\
\sigma_{NEUT} & =1.05\\
\sigma_{GENIE} & =.954\\
\sigma_{NuWro} & =.911
\end{split}
\end{equation}

A data release has been  provided\cite{DataRel-TN328} that contains the double-differential cross section central values and 
associated relative covariance matrix for both the regularized and unregularized fits.

\subsection{Comparisons to Models.}

\section{Discussion and Summary}

We have performed a measurement of the $\bar \nu_\mu$ CC 
double differential cross section on 
water without pions in the final state averaged over the T2K antineutrino beam flux. 
The measurement method in momentum-$\cos \theta$ bins included a 
likelihood fit with unfolding to correct for bin to bin smearing. The data was fit without regularization and
with regularization to reduce bin to bin fluctuations that are possible when using unfolding methods.
The regularized and unregularized results were nearly identical. 
The comparisons with the NEUT, GENIE,
 and NuWro models find a lowest $\chi^2$ for
NuWro where nearly all of the 19 measured data bins agreed within 1 standard deviation of the NuWro 
predictions. 

In summary, the first measurements of antineutrino cross sections on water
were presented and are found to be 
in agreement with several MC model predictions including NEUT,  
which is extensively used in the T2K measurements of antineutrino interactions at the SuperK far detector.  
These antineutrino measurements 
and comparisons to Monte Carlo predictions are extremely 
important for the measurements of  
the antineutrino oscillation rates and the search for CP violation
by T2K and for the development of future
long baseline neutrino experiments. 

\section{Acknowledgements}

We thank the J-PARC accelerator team for the superb accelerator performance
and CERN NA61/SHINE colleagues for providing particle production data and
for their collaboration. We acknowledge the support of MEXT, Japan;
NSERC, NRC and CFI, Canada; CEA and CNRS/IN2P3, France; DFG, Germany;
INFN, Italy; Ministry of Science and Higher Education, Poland; RAS,
RFBR and the Ministry of Education and Science of the Russian Federation;
MEST and NRF, South Korea; MICINN and CPAN, Spain; SNSF and SER, Switzerland;
STFC, U.K.; \%NSF and DOE, U.S.A. We also thank CERN for donation
of the UA1/NOMAD magnet and DESY for the HERA-B magnet mover system.
In addition, participation of individual researchers and institutions
in T2K has been further supported by funds from: ERC (FP7), EU; JSPS
and the National Institute of Informatics for SINET4 network support,
Japan; Royal Society, UK; DOE Early Career program, and the A. P.
Sloan Foundation, U.S.A. Computations were performed on the supercomputers
at the SciNet HPC Consortium. SciNet is funded by: the Canada Foundation
for Innovation (Compute Canada); the Government of Ontario; Ontario
Research Fund (Research Excellence); and the University of Toronto.

\bibliographystyle{apsrev4-1}
\bibliography{savedrecs,main}

\end{document}